\documentclass[aps,prr,superscriptaddress, twocolumn, amsmath, amssymb, notitlepage,longbibliography]{revtex4-2} 
\usepackage{float,graphicx,graphics,epsfig,subfigure,times,bm,bbm,amssymb,amsmath,amsfonts,amsthm,mathrsfs,MnSymbol}
\usepackage{graphicx} 
\usepackage{amsmath,amssymb}
\usepackage{verbatim}
\usepackage{comment}
\usepackage{mathtools}
\usepackage{dsfont}
\usepackage{xcolor}
\usepackage{algorithm2e}
\usepackage{graphicx}
\usepackage{subcaption}
\usepackage{float}
\usepackage{nicefrac}

\usepackage{thmtools}
\usepackage{thm-restate}

\usepackage{mathrsfs}
\usepackage{nccmath}
\usepackage{placeins}
\usepackage{physics}
\usepackage{comment}
\usepackage{qcircuit}
\usepackage{pgffor}
\usepackage[colorlinks,linkcolor=blue,urlcolor=blue,citecolor=blue]{hyperref}
\usepackage{cleveref}

\newcommand{\bea}{\begin{eqnarray}}
\newcommand{\eea}{\end{eqnarray}}
\newcommand{\unit}{1\!\!1}

\begin{document}
\title{Impossibility of Refrigeration and Engine Operation in Minimal Qubit Repeated-Interaction Models} 
\author{Gabrielle Barsky-Giles*}
\affiliation{Department of Physics, University of Toronto, 60 Saint George St, Toronto, Ontario M5S 1A7 Canada}
\affiliation{Department of Physics, Engineering Physics, and Astronomy, Queen’s University, Kingston ON, K7L 3N6, Canada }
\altaffiliation{These authors contributed equally to this work.}
\author{Alessandro Prositto*}
\affiliation{Department of Physics, University of Toronto, 60 Saint George St, Toronto, Ontario M5S 1A7 Canada}
\altaffiliation{These authors contributed equally to this work.}
\author{Matthew Gerry}
\affiliation{Department of Physics, University of Toronto, 60 Saint George St, Toronto, Ontario M5S 1A7 Canada}
\author{Dvira Segal}
\affiliation{Department of Chemistry and Centre for Quantum Information and Quantum Control, University of Toronto, 80 Saint George St., Toronto, Ontario M5S 3H6, Canada}
\affiliation{Department of Physics, University of Toronto, 60 Saint George St, Toronto, Ontario M5S 1A7 Canada}

\begin{abstract}
We investigate the operation of a qubit as a quantum thermal device within the repeated interaction framework, allowing for strong system-bath coupling and finite interaction times. We analyze two minimal models: an alternating-coupling setup, in which the qubit sequentially interacts with hot and cold baths, and a simultaneous-coupling setup, where both baths interact with the qubit during each collision.
For the alternating model, we obtain an exact analytical solution for the limit-cycle state, valid for arbitrary coupling strengths and collision durations. Using this solution, we rigorously prove a no-go theorem for quantum refrigeration. We further demonstrate that, although work can be generated locally at individual system-bath contacts, the total work over a cycle is always nonpositive, precluding engine operation. In the absence of work, the model describes pure heat conduction, for which we derive a closed-form  expression for the heat current and show that it exhibits a nonmonotonic turnover behavior.
The simultaneous-coupling model is analyzed perturbatively. In the short-collision-time limit, it reproduces the same steady-state behavior as the alternating model, reinforcing the generality of the constraints identified. 
Our results establish fundamental limitations on qubit-based quantum thermal machines operating under Markovian repeated interactions and highlight the need for enriched models to realize functional quantum thermal devices.
\end{abstract}
\date{\today}

\maketitle

\section{Introduction}
The ability to engineer efficient nanoscale thermal machines represents a fundamental milestone for quantum information processing and nanotechnology, particularly in enabling the control of near-term quantum devices \cite{Nayem2025,Auffeves2022}. Consequently, significant efforts have been devoted to the practical realization of such machines, which may allow cooling beyond the limits imposed by simple thermalization with a single reservoir, or conversion of otherwise wasted heat into useful work \cite{Dutta2021,Cleverson2025,Karen2025}.
Experimental efforts have included the realization of thermal machines with quantum working fluids based on trapped ions \cite{Maslennikov2019} and atoms \cite{Rossnagel2016}, semiconductor quantum dots \cite{Josefsson2018}, and superconducting qubits   \cite{Aamir2025}. However, fundamental questions remain unanswered. %
%
Quantum thermal machines can be made more efficient than their analogous classical heat engines. This higher efficiency can be achieved in different ways, including leveraging steady-state coherence \cite{Poem2019,RomanAncheyta2021,Manzano2021, Manzano2025,Petruccione2019,DeChiara2022}, squeezed thermal states \cite{Klaers2017,Lutz2014,Manzano2016,Yi2012}, asymmetric couplings to thermal baths \cite{Ahmadi2025}, and many-body cooperative effects \cite{Kurziki2018QuantumAdvantage,DelCampo2016,Ghosh2023,Divakaran2021}.
However, these quantum advantages are achieved only in specific working regimes depending on the particular type of quantum thermal machine considered. Outside these regimes, the same effects providing quantum advantages can become detrimental; see, for example, Refs. \citenum{Manzano2021,Manzano2025}.
Identifying these regimes, and the mechanisms that enable them, is therefore essential for optimizing the operating conditions of quantum thermal machines and leveraging quantum effects to achieve performance beyond classical limits. 

Theoretical descriptions of open quantum systems--including quantum thermal machines--typically rely on Markovian quantum master equations derived under {\it weak-coupling} and secular approximations \cite{Petruccione2002,Martin2010, Breuer2022}. In particular, thermal machines consisting of periodic interactions with a hot and a cold reservoir are commonly modeled via the Lindblad quantum master equation \cite{Dutta2021,Arrachea2025,Brunner2017,Mauro2018,Brunner2024,Manzano2025,Manzano2021}. This approach inherently makes strong assumptions about the character of the reservoirs (Markovian, stationary, Gaussian), the system-bath interaction (weak-coupling),  and often eliminates system–bath coherences from the evolution. 

The repeated interaction (RI) formalism (also known as \textit{collision model}) offers an alternative framework to describe open quantum systems. In this approach, system–bath interactions are treated microscopically as sequences of unitary collisions with environmental ancillas, see for example \cite{Rau1963,Palma2022, Merkli2014,Pocrnic2025,Campbell2021,Buzek2005, Poletti2023,Buzek2012,Giovannetti2013,Strunz2016NonMarkovian}. In the simplest RI scenario, these ancillas are considered identical, uncorrelated, and noninteracting. Each ancilla can interact with the system only once before it is discarded or, equivalently, refreshed to its initial state. Crucially, this scenario describes a Markovian open quantum system dynamics without making any assumptions about the system-bath coupling strength, thus allowing for a distinction to be made between Markovianity and strong-coupling (see, for instance, Ref.~\citenum{Prositto2026}).  

Due to their simplicity and versatility, repeated-interaction (RI) models have made substantial contributions to the development of the theoretical framework of quantum thermodynamics \cite{Barra2015,Palma2022,Zambrini2022,Parrondo2022}.
In particular, RI models have been employed to extend quantum thermodynamics beyond the weak-coupling regime \cite{Strasberg2019}, to resolve thermodynamic inconsistencies arising in local master equations \cite{Mauro2018}, and to investigate the thermodynamic properties of structured systems \cite{Xia2022}. They have also provided valuable insights into the thermodynamic role of quantum coherences and system–environment correlations \cite{Landi2019,Xia2019}, as well as an intuitive bridge between quantum thermodynamics and quantum information processing and computation \cite{Pocrnic2025,Brandes2017,Pautrat2017}. In addition, RI models have been shown to be well suited for applications in quantum thermometry \cite{Landi2019Thermometry,Gerasimov2025} and quantum control theory \cite{Kurizki2008}. Finally, in Refs.~\cite{Prositto2025Dynamics,RamonEscandel2025,Kurizki2019}, RI models were employed to study thermalization dynamics in simple quantum systems, providing insights relevant to the design of efficient algorithms for preparing thermal states on near-term digital quantum computers.

Moreover, owing to their ability to disentangle non-Markovian effects from strong coupling and to provide access to dynamical regimes beyond the Lindblad limit, while often allowing for analytic solutions, RI models are particularly well suited for probing the operational limits of quantum thermal machines, especially stroke-based architectures, and for assessing the role of genuinely quantum resources in their performance \cite{Landi2020,DeChiara2020,Bellomo2021,Kosloff2014Machine}. In this context, RI models have enabled systematic investigations of the impact of quantum coherences \cite{RomanAncheyta2021,DeChiara2022,DeChiara2024,Juan2019,DeChiara2022}, quantum and classical correlations \cite{Mauro2020}, the form of the system–bath interaction \cite{Xia2022ThermalMachines}, and quantum measurements \cite{Goold2023ThermalMachines} on the performance of quantum thermal machines. They have also been used to discuss fundamental limits on efficiency at the nanoscale and their dependence on the properties of heat baths \cite{Wehner2019}. Finally, experimental implementations of quantum thermal machines inspired by the RI paradigm have been realized using micromasers \cite{Scully2003} and ultracold atomic gases \cite{Widera2021}.

An analytically exact solution for the general RI dynamics of a quantum thermal machine is, however, still lacking, and even in conceptually simple settings such as all-qubit models. Indeed, most of the existing literature is restricted either to the weak-coupling regime or to the short-collision-time (stroboscopic) limit, where an effective quantum master equation can be straightforwardly derived; thermodynamic quantities are then typically evaluated within this framework using the definitions introduced in Ref.~\citenum{Barra2015}. To avoid complications arising from explicit time dependence in the master equation, studies often focus on quantum thermal machines in which heat exchange and work production occur in two distinct steps, or strokes, as in the Otto cycle. Moreover, when a working fluid is considered, it is usually modeled as a system more complex than a single qubit.

Obtaining analytically exact solutions for the dynamics of minimal models, such as an all-qubit setup with a simple working cycle in which heat and work are exchanged simultaneously and the working fluid consists of a single qubit, would be highly valuable, not only for elucidating the fundamental mechanisms underlying the operation of quantum thermal machines, but also for providing reliable benchmarks beyond perturbative regimes. This is the objective of the present work.

\begin{figure}[htbp]
    \centering
    \includegraphics[width=0.9\linewidth,trim={0cm 0.1cm 0cm 0cm}, clip]{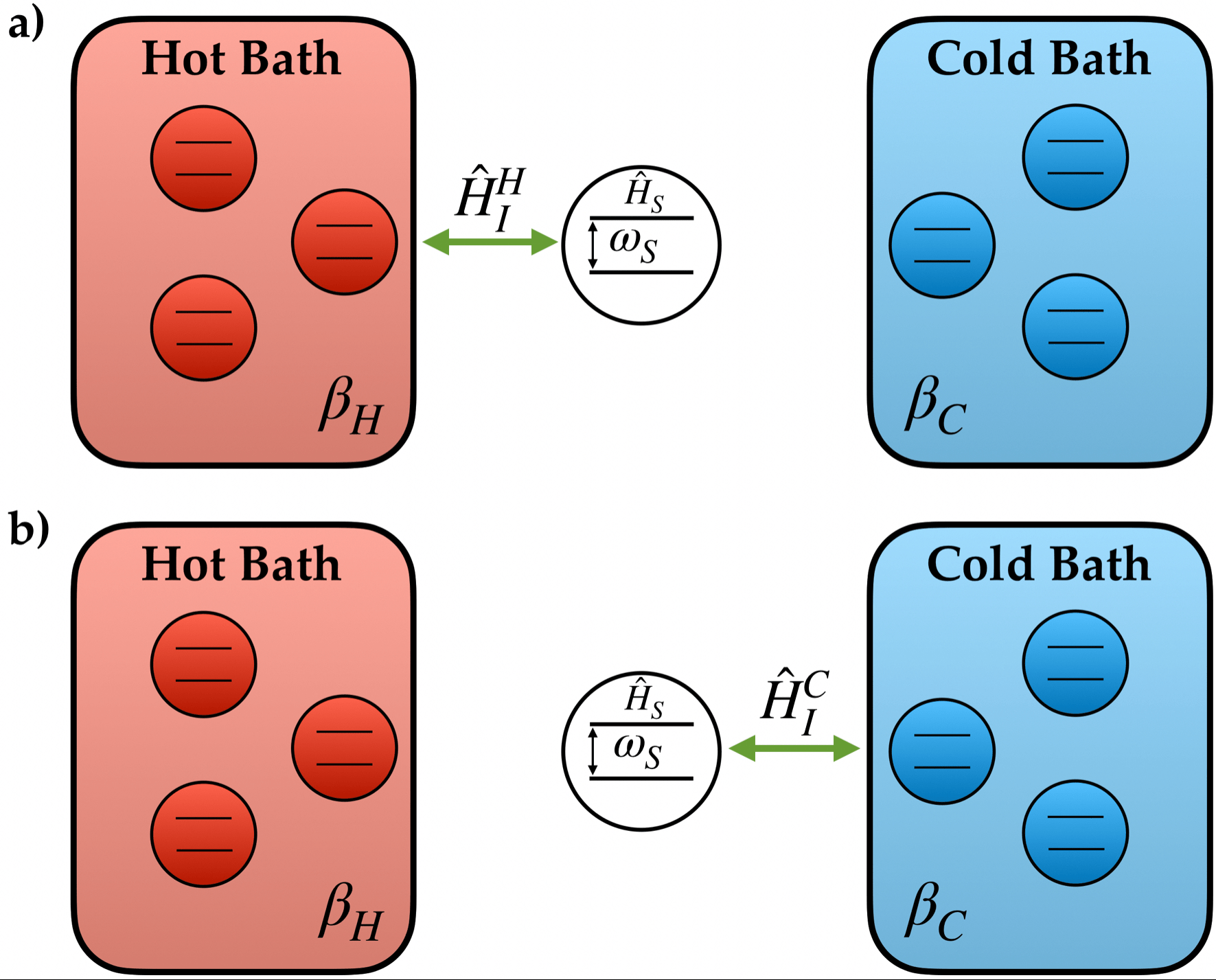}
    \caption{Alternating model for a quantum thermal machine. A two level system ($\hat H_S$) collides with each bath separately, alternating between collisions with (a) hot and (b) cold ancillas.}
    \label{fig:S1}
\end{figure}

In this work, we investigate whether a qubit can operate as a quantum thermal machine within the simplest RI framework, considering two scenarios: alternating coupling to hot and cold baths, and simultaneous coupling to both baths during each collision. For schematic representations of the two models, see Fig. \ref{fig:S1} and Fig. \ref{fig:SketchS}, respectively. We address the following questions:
(i) For a purely heat-conducting setup, what are the system’s thermal transport properties, and what signatures signal the strong-coupling regime?
(ii) Can the system operate as a quantum thermal refrigerator?
(iii) Can it function as a heat engine, converting heat into work?

The alternating model is solved exactly and analytically, whereas the simultaneous model is treated perturbatively using a Dyson-series expansion. Despite these different solution strategies, both models exhibit common features:
(i) the heat exchange displays oscillatory behavior as a function of the coupling energy, with the heat current showing a turnover behavior as a function of the system-bath coupling energy;
(ii) the system cannot operate as a quantum thermal refrigerator, as we rigorously prove for the alternating model (Sec. \ref{Sec:no_refrigerator}); simulations in the simultaneous-coupling model yield consistent conclusions;
(iii) simulations show that while work can be generated locally at one of the contacts, the total work contribution is nonpositive. That is, in total, work is performed on the system and is subsequently dissipated as heat, meaning that the system cannot function as a heat engine. 
Together, these results establish fundamental constraints on qubit-based thermal machines within a Markovian RI dynamics. 

The paper is organized as follows: In Sec. \ref{sec:Alter} we focus on a quantum thermal machine with alternating coupling to heat baths, studying its performance and limitations as a quantum thermal device. 
We solve exactly the limit cycle state of the model and prove a ``no-go theorem" for quantum refrigeration.
In Sec. \ref{sec:Simult} we solve the simultaneous model perturbatively, at weak system-ancilla couplings. We summarize our work in Sec. \ref{sec:Summ}. 
Throughout this study, we work in units of $\hbar\equiv1$ and $k_B\equiv 1$.

\section{Alternating Coupling Thermal Machine} 
\label{sec:Alter}

In this section, we focus on the alternating mode of operation, in which a qubit system, 
the working fluid of a quantum thermal machine,
is sequentially coupled to either a cold or a hot bath. Throughout this study, we assume Markovian dynamics and model each bath as a collection of independent, uncorrelated, and noninteracting ancillary qubits, all prepared in a Gibbs thermal state at inverse temperature $\beta_H$ ($\beta_C$) for the hot (cold) bath.
Each ancilla is discarded after the interaction or, equivalently, refreshed to its initial thermal state.
Although the collision dynamics is Markovian, each collision may be strong and long, thus resulting in an overall dynamics that deviates from the Lindblad quantum master equation (QME) limit \cite{Prositto2025Dynamics}. For a schematic representation, see Fig. \ref{fig:S1}.

Within the repeated-interaction framework, the system-ancilla interaction is repeatedly switched on and off. The energy associated with this process is identified as ``work", as it originates from a time-dependent modulation of the interaction Hamiltonian rather than energy exchanged with a thermal reservoir \cite{Barra2015}. In contrast, heat is defined as the energy transferred to the ancilla during its interaction with the qubit: each ancilla is initially prepared in a thermal Gibbs state and subsequently discarded, so that the change in the internal energy of the ancilla corresponds to heat exchanged between the system qubit and the ancilla itself. 
This distinction enables a consistent thermodynamic interpretation of energy flow in repeated interaction models, even beyond the weak-coupling regime. In particular, for partial-swap operations and resonant system-ancilla frequencies, the work contribution vanishes, and the model describes a purely heat transport setup  \cite{Prositto2025Dynamics}. 
The concrete definitions of work and heat are provided in Sec.~\ref{sec:Alter-thermo}.

We now introduce the Hamiltonian of the thermal machine with alternating coupling to the two baths.
The Hamiltonian for the system ($S$), the hot ancilla ($H$), and the cold ancilla ($C$) are given by
\begin{equation}
 \hat{H}_S = - \frac{\omega_S}{2}\hat{\sigma}_z^S, \,\,\, \hat{H}_H = - \frac{\omega_H}{2}\hat{\sigma}_z^H, \,\,\, \hat{H}_C = - \frac{\omega_C}{2}\hat{\sigma}_z^C,
\end{equation}
where $\omega_{\alpha}$, $\alpha=S,H,C$, represents the energy splitting of the system-qubit (S) and the two baths' ancilla-qubits, hot (H) and cold (C). We indicate by $\hat{\sigma}_{m}^{\alpha}
,\:m=x,y,z$, the Pauli operator along the $m$-direction in the Hilbert space of the 
$\alpha=S,H,C$ qubit.
In what follows, we typically assume that the ancilla qubits are all resonant with the system, thus $\omega_{\alpha} $ are all identical. However, some of our derivations are done in more general settings, beyond the resonant restriction. 

The system-ancilla interaction Hamiltonian is defined for collisions with either the hot ($\hat{H}_{I}^{H}$) or cold ($\hat{H}_{I}^{C}$) ancilla as
\begin{align}
    \hat{H}_I^H &= J_{xx}^H\hat{\sigma}_x^S \otimes \hat{\sigma}_x^H + J_{yy}^H\hat{\sigma}_y^S \otimes \hat{\sigma}_y^H,\nonumber\\[0.3em]
    \hat{H}_I^C &= J_{xx}^C\hat{\sigma}_x^S \otimes \hat{\sigma}_x^C + J_{yy}^C\hat{\sigma}_y^S \otimes \hat{\sigma}_y^C . 
\end{align}
We assume that collisions occur frequently, with the time interval between successive interactions being sufficiently short that any ``down time'' without collisions can be neglected.

In this work, we analyze the function of the system as a thermal machine as we vary the four interaction parameters, $J_{xx}^{C}$, $J_{yy}^{C}$, $J_{xx}^{H}$, and $J_{yy}^{H}$, as well as the collision duration, $\tau$, which for simplicity is assumed equal for all collisions.  
Since in general $[\hat{H}_{I}^{A}, \hat{H}_{S}] \neq 0$ for $A = H, C$, the interaction energy evolves in time and, as discussed above, can be interpreted as a contribution of work to the overall energy balance. 
When $J_{xx} = J_{yy}$, the interaction reduces to a partial-swap operation. Together with the resonance condition on frequencies, these are sufficient conditions for the work component to vanish.

For each bath, the ancillas are prepared in a Gibbs thermal state at inverse temperature $\beta_A$ with $A$=$H$, $C$, namely,
\begin{equation}
    \rho_{A} = 
    \begin{pmatrix}
        p_{A}&c_{A}\\[0.5em](c_{A})^{*}&1-p_{A}
    \end{pmatrix} = 
    \begin{pmatrix}
        \frac{1}{1+e^{-\beta_A\omega_{A}}}&0\\[0.5em]0&\frac{e^{-\beta_A\omega_{A}}}{1+e^{-\beta_A\omega_{A}}}
    \end{pmatrix}.
\end{equation}
Here, $p_{A}$ and $c_{A}=0$ represent the population in the ground state and the coherence of the ancilla, respectively, upon reset.

The initial state of the system, $\rho_{S}^{(0)}$, can be taken randomly. In the $n^{th}$ RI step, the density matrix that describes the state of the system, $\rho_{S}^{(n)}$, takes the form 
\begin{equation}
    \rho_S^{(n)} = 
    \begin{pmatrix}
        p_S^{(n)} & c_S^{(n)} \\[0.5em] 
        ( c_S^{(n)})^* & 1- p_S^{(n)}
    \end{pmatrix}.
\end{equation}
In the next subsections, we study the system relaxation dynamics and its state in the limit cycle (steady state), indicated by
\begin{equation}
    \rho_S^{(\infty)} = 
    \begin{pmatrix}
        p_S^{(\infty)} & c_S^{(\infty)} \\[0.5em]
        ( c_S^{(\infty)})^* & 1- p_S^{(\infty)}
    \end{pmatrix}.
\end{equation}
%

\subsection{Population dynamics and the steady state: single bath}
\label{sec:1bath}

Before describing the dynamics of the alternating-coupling (two baths) model, we briefly summarize the single-bath scenario, in which the system undergoes repeated collisions with ancillas drawn from the same bath \cite{Prositto2025Dynamics}. In this case, the total Hamiltonian describing a single collision is given by
\bea
    \hat{H}_{tot} &= \hat{H}_I+ \hat{H}_S \otimes \unit_A + \unit_S  \otimes\hat{H}_A,
\eea 
where the symbol $\unit_{\alpha}$ with $\alpha=S,A$ indicates the identity operator in the Hilbert space of the system or the ancilla. A single RI step evolves the state of the system according to
\begin{equation}
    \rho_S^{(n+1)} = \text{Tr}_A \Big[\hat{U}(\tau)(\rho_S^{(n)} \otimes \rho_{A})\hat{U}^{\dagger}(\tau)\Big],
    \label{eq:RI_step}
\end{equation}
with the time evolution operator $\hat{U}(\tau) = e^{-i \hat{H}_{tot} \tau}$.
As shown in Ref.~\citenum{Prositto2025Dynamics}, simplifying Eq.~(\ref{eq:RI_step}) yields the following evolution equation for the system ground-state population:
\bea
p_S^{(n+1)} = \left(1-\kappa_{\theta}-\kappa_{\phi}\right)p_S^{(n)} +
\left(\kappa_{\theta} + \kappa_{\phi}\right) p_S^{(\infty)},
\label{eq:col}
\eea
with coefficients
\bea
\kappa_{\theta} &=& 4\frac{(J_{xx}+J_{yy})^2}{\theta^2}\sin^2{\left(\frac{\theta\tau}{2}\right)},
\nonumber\\
\kappa_{\phi} &=& 4\frac{(J_{xx}-J_{yy})^2}{\phi^2}\sin^2{\left(\frac{\phi\tau}{2}\right)},
\label{eq:kappatp}
\eea
and frequencies 
\bea
\theta&=&\sqrt{4(J_{xx}+J_{yy})^2 + (\omega_A-\omega_S)^2},
\nonumber\\
\phi&=&\sqrt{4(J_{xx}-J_{yy})^2 + (\omega_A+\omega_S)^2}.
\eea
Here, $\omega_A$ ($\omega_S$) is the energy gap of the ancilla (system), potentially different. 

As for the steady state of a qubit repeatedly colliding with ancillas of the same bath,  it is given by \cite{Prositto2025Dynamics} 
\bea
p_S^{(\infty)}= \frac{\kappa_{\theta} p_A + \kappa_{\phi}(1-p_A)}{\kappa_{\theta}  + \kappa_{\phi}},
\label{eq:psinf1}
\eea
which makes it clear that only when $J_{xx}=J_{yy}$, the system reaches the (canonical) state of the bath.
Using Eq. (\ref{eq:psinf1}), we rewrite Eq. (\ref{eq:col}) as
\bea
p_S^{(n+1)} = \left(1-\kappa_{\theta}-\kappa_{\phi}\right)p_S^{(n)} +
\left[\kappa_{\theta}p_A + \kappa_{\phi}(1-p_A)\right].
\nonumber\\
\label{eq:coln}
\eea
Since the map (\ref{eq:RI_step}) is Markovian, it is not surprising that the population dynamics follows a Markov process, which is dictated by the coefficient
$1-\kappa_{\theta}-\kappa_{\phi}$. 
As demonstrated in Ref.~\citenum{Prositto2025Dynamics}, the map (\ref{eq:RI_step}) leads to decoupled evolution of coherences and populations, with coherences vanishing in the steady state.


\subsection{Population dynamics and the limit cycle: two alternating baths}

We proceed to the two-bath scenario. By generalizing the single-bath solution, we obtain an {\it exact} analytical expression for the system ground-state population in the limit cycle, valid for arbitrary collision times and interaction strengths.
We now describe this generalization.

The system interacts with one bath at a time. Accordingly, the evolution during each collision is governed by one of two distinct total Hamiltonians,
\begin{align}
    \hat{H}_{tot}^H &= \hat{H}_I^H+ \hat{H}_S \otimes \unit_H + \unit_S  \otimes\hat{H}_H\nonumber,\\[0.3 em]
    \hat{H}_{tot}^C &= \hat{H}_I^C+ \hat{H}_S \otimes \unit_C + \unit_S  \otimes\hat{H}_C\:.
\end{align}
Depending on the type of ancilla ($H$ or $C$), the collision between the system and an ancilla is described by the unitary 
\bea
\hat{U}_{A}(\tau) = e^{-i \hat{H}_{tot}^{A} \tau}, \,\,\,\, A=H,C\:.
\label{eq:UA}
\eea
After each collision, the state of the system evolves according to Eq. (\ref{eq:RI_step}).
Here, we make a distinction between the two propagators, $\hat{U}_H(\tau)$ and $\hat{U}_C(\tau)$, since the interaction parameters with the different baths may differ; in the special case where $J_{xx}^H=J_{xx}^C$ and $J_{yy}^H=J_{yy}^C$, $\hat{U}_H(\tau)=\hat{U}_C(\tau)$; the difference between the ancillas'  temperatures is reflected in the state at the beginning of each collision, with the ancilla state alternating between $\rho_H$ and $\rho_C$.

Applying the single-bath result, Sec. \ref{sec:1bath}, to the alternating bath scenario, we note that after a single collision with either the hot or the cold bath, $A = H, C$, the system evolves according to Eq.~(\ref{eq:coln}),
\bea
p_S^{(n+1)} = g_A p_S^{(n)} + f(p_A),
\label{eq:colHCs}
\eea
where
\bea 
g_A=\left(1-\kappa_{\theta}^A-\kappa_{\phi}^A\right), \,\,\,\,
f(p_A)=\kappa_{\theta}^Ap_A + \kappa_{\phi}^A(1-p_A), 
\nonumber\\
\eea
with coefficients
\bea
\kappa_{\theta}^A &=& 4\frac{(J_{xx}^A+J_{yy}^A)^2}{\theta_A^2}\sin^2{\left(\frac{\theta_A\tau}{2}\right)}
\nonumber,\\[0.3 em]
\kappa_{\phi}^A &=& 4\frac{(J_{xx}^A-J_{yy}^A)^2}{\phi_A^2}\sin^2{\left(\frac{\phi_
A\tau}{2}\right)},
\label{eq:kappa}
\eea
and frequencies
\bea
\theta_A&=&\sqrt{4(J_{xx}^A+J_{yy}^A)^2 + (\omega_A-\omega_S)^2}
\nonumber,\\
\phi_A&=&\sqrt{4(J_{xx}^A-J_{yy}^A)^2 + (\omega_A+\omega_S)^2\:}.
\label{eq:freq}
\eea
A cycle, however, consists two collisions: one with a hot ancilla and one with a cold ancilla. 
We therefore construct a recursive relation for the system ground-state population after a complete cycle by applying Eq.~(\ref{eq:colHCs}) twice. 
Since the resulting expression depends on whether the cycle is defined as a hot collision followed by a cold one, or vice versa, we obtain
\bea
\label{nsecond}
    p_S^{(n+2)}=
    \left\{
     \begin{array}{@{}l@{\thinspace}l}
       f(p_C)+g_C f(p_H)+g_C g_H p_S^{(n)}  
         \text{\,\,\,\,  (hot, cold)} \\[0.5 em]
       f(p_H)+g_H f(p_C)+g_H g_C p_S^{(n)}  
         \text{\,\,\,\, (cold, hot) } \\
     \end{array}.
   \right.
   \nonumber\\
\eea
Let's assume now that a cycle is defined as a collision with a hot ancilla followed by a cold one [Eq.~(\ref{nsecond}), first line]. We denote by  $p_S^{(\infty, C)}$ the ground-state population in the limit cycle, where the superscript $C$ indicates that this population is evaluated immediately after the collision with the cold ancilla.
According to Eq.~(\ref{eq:colHCs}), the population decays geometrically towards its steady state value with a coefficient $g$ that we can immediately identify,
\begin{equation}
    p_S^{(n+2)}-p_S^{(\infty,C)}=g\left( p_S^{(n)}-p_S^{(\infty,C)}\right).
\end{equation}
Rearranging, we get,
\begin{equation}
    p_S^{(n+2)}=g p_S^{(n)}+\left(p_S^{(\infty,C)}-g p_S^{(\infty,C)}\right).
\end{equation}
Noting that the right hand side must be equal to that of Eq. (\ref{nsecond}) (first case), we identify $g= g_C g_H$, and further, we resolve the limit-cycle population,
\begin{equation}
    p_S^{(\infty,C)} = \frac{f(p_C)+g_C f(p_H)}{1-g_C g_H}.
\end{equation}
Using the same procedure, we obtain the population of the ground-state of the system in the limit cycle, immediately after a collision with a hot ancilla [Eq.~(\ref{nsecond}), second line],
\begin{equation}
    p_S^{(\infty,H)} = \frac{f(p_H)+g_H f(p_C)}{1-g_C g_H}.
\end{equation}
These limit-cycle populations can equivalently be expressed in terms of variables defined in Eqs.~(\ref{eq:kappa})-(\ref{eq:freq}), giving
\begin{widetext}
\bea
    p_S^{(\infty,C)} &=& \frac{\kappa_{\theta}^C p_C+\kappa_{\phi}^C (1-p_C)+(1-\kappa_{\theta}^C-\kappa_{\phi}^C) [\kappa_{\theta}^H p_H+\kappa_{\phi}^H (1-p_H)]}{1-(1-\kappa_{\theta}^C-\kappa_{\phi}^C) (1-\kappa_{\theta}^H-\kappa_{\phi}^H)},
\nonumber\\[0.3em]
    p_S^{(\infty,H)} &=& \frac{\kappa_{\theta}^H p_H+\kappa_{\phi}^H (1-p_H)+(1-\kappa_{\theta}^H-\kappa_{\phi}^H) [\kappa_{\theta}^C p_C+\kappa_{\phi}^C (1-p_C)]}{1-(1-\kappa_{\theta}^C-\kappa_{\phi}^C) (1-\kappa_{\theta}^H-\kappa_{\phi}^H)}.
\label{eq:pop}
\eea
\end{widetext}
As for coherences, they decay to zero in the limit cycle, similarly to Ref.~\citenum{Prositto2025Dynamics}.

The limit cycle solution (\ref{eq:pop}) of the system constitutes one of the main results of our work: an exact closed-form expression for the population of the system in the limit cycle, valid for arbitrary coupling strength and duration of interaction.

It is possible to establish certain bounds on limit-cycle populations. Recall that
$p_{H,C}$ refers to the population of the ancillas in the ground state before a collision.
Using the relations $p_H \leq p_C$, $p_C \geq 1/2$, and $p_H + p_C \geq 1$, one can readily prove from Eq. (\ref{eq:pop}) that 
\bea
p_S^{(\infty, C)} \leq p_C \quad \text{and} \quad p_S^{(\infty, H)} \leq p_C,
\label{eq:boundpop}
\eea
see Appendix~\ref{AppA}. 
These inequalities imply that, for any choice of interaction parameters, collision times, and temperatures, the system's ground-state population is always less than or equal to that of the cold ancilla.  

Translating system's populations into an effective temperature defined for the system after a collision with an A-type ancilla, in the limit cycle, we get
\[
\beta_S^A = -\frac{1}{\omega_S} \ln\left( \frac{1 - p_S^{(\infty,A)}}{p_S^{(\infty,A)}}
\right).\] 
The relation $p_S^{(\infty, A)}\leq p_C$ can now be restated as follows: The effective temperature of the system cannot achieve a value {\it lower} than that of the cold ancillas, neither before nor after a collision with them. 
This implies that the system cannot extract heat from the cold bath and, therefore, cannot operate as a quantum refrigerator. 
A direct proof of this impossibility is provided in Sec.~\ref{Sec:no_refrigerator}, where we analyze heat exchange in detail in the alternating coupling model.

\begin{figure*}[ht!]
    \centering
    \includegraphics[width=1\linewidth]{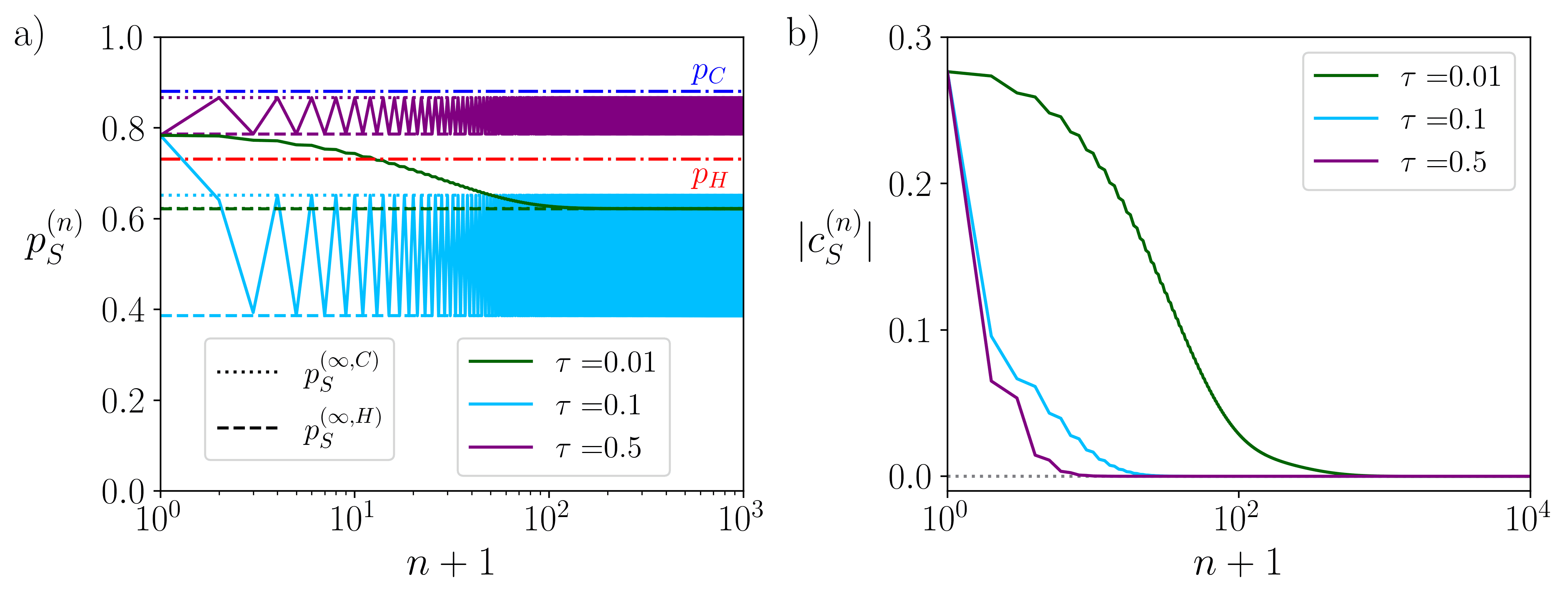}
\caption{System relaxation dynamics in the alternating-coupling model from a certain initial state toward the limit cycle with $n$ as the number of collisions.  
(a) Relaxation of the system's ground-state population to limit cycle values, alternating between $p_S^{(\infty,C)}$ and $p_S^{(\infty,H)}$. 
As a reference, the populations of ancillas from the baths are shown as dash-dotted lines, blue for the cold bath, 
$p_C$, and red for the hot bath, $p_H$.
(b) Decoherence dynamics in the qubit system. 
We consider asymmetric coupling strengths to the baths, $J^{H}_{xx} =4$, $J^{H}_{yy} = 16$ and $J^{C}_{xx} = 2$, $J^{C}_{yy} = 8$.  The system and ancilla's frequencies are set to $\omega_{S} = \omega_{C} = \omega_{H} = 1$, and the cold (hot) bath inverse temperatures are $\beta_{C} = 2$ ($\beta_{H} = 1$).} 
    \label{fig:popA}
\end{figure*}
%

\begin{figure*}
    \centering
    \includegraphics[width=1.0\linewidth]{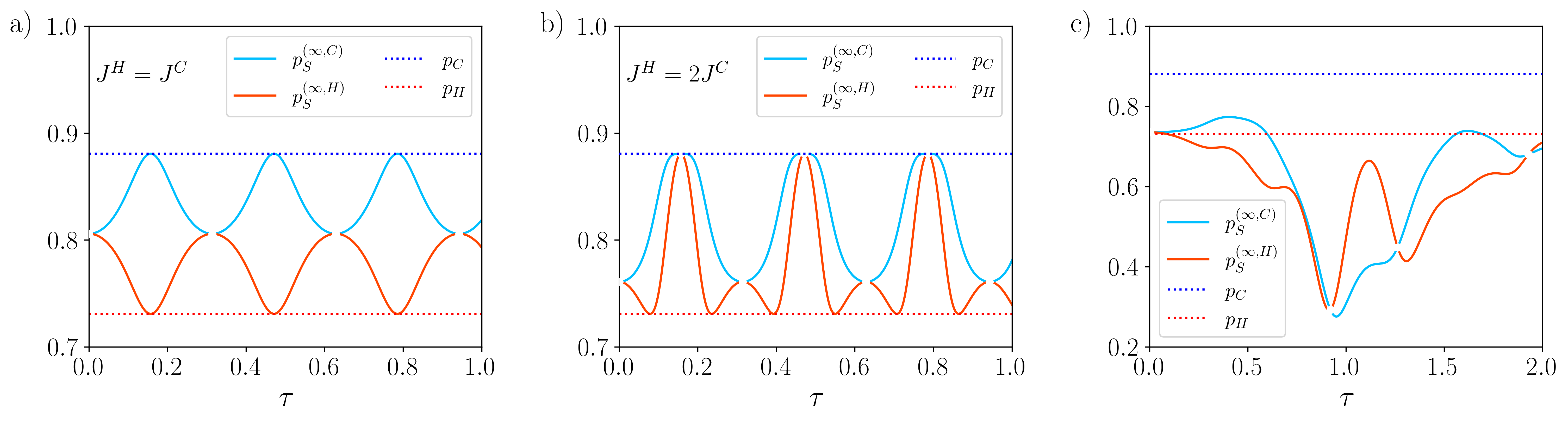}
\caption{Limit cycle of the alternating-coupling model: Ground-state population of the qubit system as a function of the collision time $\tau$. 
(a) Symmetric coupling, with $J_{xx}^{C} = J_{yy}^{C} = 5$ and $J_{xx}^{H} = J_{yy}^{H} = 5$. 
(b) Asymmetric coupling, with $J_{xx}^{C} = J_{yy}^{C} = 5$ and $J_{xx}^{H} = J_{yy}^{H} = 10$. (c) General coupling, with $J_{xx}^{C} =1$, $J_{yy}^{C} = 2$, $J_{xx}^{H}=4$ , and $J_{yy}^{H} = 3$. 
Other parameters are $\omega_{S} = \omega_{C} = \omega_{H} = 1$,
$\beta_{C} = 2$ and $\beta_{H} = 1$. 
For reference, the ground-state populations of the cold and the hot bath ancillas are indicated by blue (cold, $p_C$) and red (hot, $p_H$) dotted lines.}
\label{fig:figpss}
\end{figure*}


It is instructive to consider special limits to Eq. (\ref{eq:pop}) and to build intuitions on them. In the case where all four interaction energies are identical, equal to $J$, and the system and ancilla's frequencies are equal, $\omega_S=\omega_A$, ($A=H,C$), we have
$\kappa_{\theta}^A =  \sin ^2\left(2 J \tau \right)$, 
$\kappa_{\phi}^A=0$.
Substituting these simplifications into Eq.~(\ref{eq:pop}), we find that
\bea
    p_S^{(\infty,C)} &=& \frac{p_H \cos (4 J \tau)+2 p_C+p_H}{\cos (4 J \tau)+3},
\nonumber\\[0.3em]
    p_S^{(\infty,H)} &=& \frac{ p_C \cos (4 J\tau)+ p_C+2 p_H}{\cos (4 J\tau)+3}.
    \label{eq:popJ}
\eea
As a sanity check, we note that the limit-cycle populations satisfy $p_S^{(\infty,C)}\geq p_S^{(\infty,H)}$, that is, the system's state immediately after a collision with the cold ancilla is always effectively colder than after a collision with a hot ancilla. 
Further, assuming $J \tau \ll 1$ and expanding in Taylor series to the third order, the limit cycle populations become
\bea
p_S^{(\infty,C)} &=& \frac{p_C+p_H}{2}+(J\tau)^2 (p_C-p_H)+O\left((J\tau)^4\right),
    \nonumber\\[0.3em]
    p_S^{(\infty,H)} &=&\frac{p_C+p_H}{2}-(J\tau)^2 (p_C-p_H)+O\left((J\tau)^4\right).\nonumber\\
\eea
In this limit, the system thermalizes to the averaged population. However, when the interaction energy is strong or $\tau$ is not short, the population of the system after each half cycle deviates from the average population, and it is closer in value to that of the bath that the system had just interacted with, see Eq. (\ref{eq:popJ}).

In the limit of $\tau\to0^{+}$, the limit cycle population in Eq. (\ref{eq:pop}) converges to
\begin{equation}
    \begin{split}
    &p_{S}^{(\infty,C)} =  
    p_{S}^{(\infty,H)}\\& =  \frac{\left(J_{xx}^{C}-J_{yy}^{C}\right)^{2}+\left(J_{xx}^{H}-J_{yy}^{H}\right)^{2} +4J_{xx}^{C}J_{yy}^{C}p_{C}+4J_{xx}^{H}J_{yy}^{H}p_{H}}{2\left[\left(J_{xx}^{H}\right)^{2}+\left(J_{yy}^{H}\right)^{2}+\left(J_{xx}^{C}\right)^{2}+\left(J_{yy}^{C}\right)^{2}\right]},
        \label{eq:popAdt}
    \end{split}
\end{equation}
In Sec. \ref{sec:Simult} we find the same result for the simultaneous model under the short collision time limit.

We illustrate the dynamics of the system from an initial state to the limit cycle in Fig.~\ref{fig:popA}(a)-(b). 
We observe that for long collision times, the limit-cycle populations $p_S^{(\infty, H)}$ and $p_S^{(\infty, C)}$  are distinct; the system population visibly alternates between two values as shown in Fig.~\ref{fig:popA}(a). In contrast, in the limit $\tau \to 0^{+}$, these populations converge, $p_S^{(\infty, H)} \to p_S^{(\infty, C)}$.
We also analyze the behavior of coherences in the model, Fig. \ref{fig:popA}(b). The coherence dynamics decays to zero over time, consistent with analytical results and observations for the single-bath model reported in Ref.~\citenum{Prositto2025Dynamics}.

We focus on the limit-cycle behavior of the model in Fig. \ref{fig:figpss}. The population of the ground-state of the system is shown both after a collision with a cold ancilla, $p_{S}^{(C,\infty)}$ and after a collision with a hot ancilla, $p_{S}^{(H,\infty)}$, plotted as a function of the duration of the collision, $\tau$.
We consider three cases. In  Fig. \ref{fig:figpss}(a) we use identical coupling at the two ends, $J^{H}=J^{C}$ where $J_{xx}^{A}=J_{yy}^{A}\equiv J^{A}$ with $A=H,C$.  In  Fig. \ref{fig:figpss}(b), we make the couplings asymmetric between the two sides, $J^{H}=2J^{C}$. Finally, in Fig. \ref{fig:figpss}(c), we consider the  general coupling case with all couplings distinct.
In the first two cases, we observe an oscillating behavior of the limit cycle populations with $\tau$, $p_{S}^{(\infty,C)}$  and $p_{S}^{(\infty,H)}$, due to the sinusoidal terms in Eq. (\ref{eq:pop}) [see also Eq. (\ref{eq:kappa})].
For symmetric coupling, Fig. \ref{fig:figpss}(a) shows that the limit cycle populations are symmetric around $(p_{C}+p_{H})/2$. In contrast, in Fig. \ref{fig:figpss}(b) we find that, since the hot bath is more strongly coupled to the system than the cold one,  an asymmetry develops in the oscillating pattern of the system after each collision, both with respect to the mean population and with $\tau$. Interestingly, in the general coupling case, Fig. \ref{fig:figpss}(c),  the system displays a counterintuitive thermal behavior: the ground-state population after a collision with a hot ancilla is {\it greater} than that after a collision with a cold ancilla. This means that an interaction with a cold ancilla can effectively {\it heat} the system more than a collision with a hot ancilla. Moreover, for a broad range of parameters, the system is hotter than the two thermal baths, including after a collision with a cold ancilla. These effects are allowed since the action of turning on and off the interaction energy involves a work exerted on the system.
%
Note that the set of points at which $p_{S}^{(\infty,H)}$ approaches $p_{S}^{(\infty,C)}$ corresponds to pathological values where the limit cycle is no longer defined ($\kappa_{\theta}^A=0$ and $\kappa_{\phi}^A=0$) and, thus, we exclude them from the plot. 

To summarize our observations thus far, we have obtained exact analytical expressions for the state of the system qubit in the limit cycle, with the ground state population after each collision given by Eq.~(\ref{eq:pop}) and the coherences vanishing. 
In addition, we have proven Eq.~(\ref{eq:boundpop}), which provides an upper bound on the system's population. We are now equipped to analyze the operation of the system as a thermal machine, conductor, refrigerator, engine, etc.

\subsection{Thermodynamical analysis in the limit cycle}
\label{sec:Alter-thermo}
We adopt definitions of work and heat as introduced and used in, e.g., Refs.~\citenum{Barra2015, Campbell20,Prositto2025Dynamics}. In a single collision of duration $\tau$, the work performed on the system or extracted from it corresponds to the change in the expectation value of the interaction Hamiltonian governing the collision. Heat exchanged between the system and the ancilla is given by the change in the expectation value of the ancilla's Hamiltonian. Recall that we prepare each ancilla in a thermal-canonical state.

\subsubsection{Work, heat, and internal energy}

The work performed in each collision is defined as 
\bea
    W^{(n+1)}_A &=& \text{Tr}[\left(\hat{U}_{A}^{\dag}(\tau)\hat{H}_I^{A}\hat{U}_{A}(\tau)-\hat{H}_I^{A}\right)(\rho_S^{(n)} \otimes \rho_A)]. \nonumber\\
\eea
It describes the change in the system-ancilla interaction energy due to a collision. A positive value for the work means that the energy stored as interaction energy {\it grows} due to the collision with the qubit system, thus extracting energy from the system. A negative value, instead, means that work is done on the system during a collision, reducing energy storage in the interaction term. After simplifications, in the limit cycle, we get \cite{Prositto2025Dynamics}
\begin{align}
    W_{C}^{(\infty)}=&\;\kappa_\theta^C(\omega_A - \omega_S)(p_S^{(\infty,H)}-p_C)\nonumber\\
    &-\kappa_\phi^C(\omega_A + \omega_S)\left[p_S^{(\infty,H)}-(1-p_C)\right],
\nonumber\\[0.3 em]
    W_{H}^{(\infty)}=&\;\kappa_\theta^H(\omega_A - \omega_S)(p_S^{(\infty,C)}-p_H)\nonumber\\
    &-\kappa_\phi^H(\omega_A + \omega_S)\left[p_S^{(\infty,C)}-(1-p_H)\right].
    \label{eq:W}
\end{align}
The total work performed on the baths by the system during a complete cycle is the combination of the above two expressions, $W^{(\infty)} = W_{C}^{(\infty)} + W_{H}^{(\infty)}$.

We define the heat exchanged between the system and the ancilla during a single collision of duration $\tau$ as
\bea
Q^{(n+1)}_C &=& \text{Tr}[\left(\hat{U}_C^{\dag}(\tau)\hat{H}_C\hat{U}_{C}(\tau)-\hat{H}_C\right)(\rho_S^{(n)} \otimes \rho_C)],
\nonumber\\[0.3em]
Q^{(n+1)}_H &=& \text{Tr}[\left(\hat{U}^{\dag}_H(\tau)\hat{H}_H\hat{U}_{H}(\tau)
-\hat{H}_H\right)(\rho_S^{(n)} \otimes \rho_H)].\nonumber\\
\eea
A positive value for the heat describes the heating of the ancilla after a collision with the system. In the limit cycle, using Eq. (\ref{eq:pop}), we get \cite{Prositto2025Dynamics}
\begin{align}
    Q_{C}^{(\infty)}=\omega_C\left[-\kappa_\theta^C(p_S^{(\infty,H)}-p_C)+\kappa_\phi^C(p_S^{(\infty,H)}-(1-p_C)\right],\nonumber\\
    \label{eq:QC}
\end{align}
and
\begin{align}
    Q_{H}^{(\infty)}=\omega_H\left[-\kappa_\theta^H(p_S^{(\infty,C)}-p_H)+\kappa_\phi^H(p_S^{(\infty,C)}-(1-p_H)\right].\nonumber\\
    \label{eq:Q}
\end{align}
We also define the change in the internal energy of the system after a collision with the cold bath as
\begin{equation}
    \Delta E_{SC}^{(n+1)} = \text{Tr}\left[\left(\hat{U}_{C}^{\dagger}(\tau)\hat{H}_{S}\hat{U}_{C}(\tau)-\hat{H}_{S}\right)\rho_{S}^{(n)}\otimes\rho_{C}\right],
\end{equation}
corresponding to   
\begin{equation}
    \Delta E_{SC}^{(n+1)} = -W_{C}^{(n+1)} - Q_{C}^{(n+1)}.
\end{equation}
Similarly, after a collision with a hot ancilla, we have
\begin{equation}
    \Delta E_{SH}^{(n+1)} = \text{Tr}\left[\left(\hat{U}_{H}^{\dagger}(\tau)\hat{H}_{S}\hat{U}_{H}(\tau)-\hat{H}_{S}\right)\rho_{S}^{(n)}\otimes\rho_{H}\right],
\end{equation}
and 
\begin{equation}
        \Delta E_{SH}^{(n+1)} = -W_{H}^{(n+1)}-Q_{H}^{(n+1)}.
\end{equation}


\subsubsection{Impossibility of quantum refrigeration}
\label{Sec:no_refrigerator}

We now prove that heat exchanged with the cold bath in the limit cycle is always non-negative, namely
\bea
Q_C^{(\infty)}\geq 0\:.
\eea
This result implies that heat extraction from the cold bath is impossible, and hence the model cannot function as a quantum thermal refrigerator.

First, since $p_S^{(\infty,H)} \le p_C$ (see Appendix~\ref{AppA}), the first term in the square brackets of Eq.~(\ref{eq:QC}), namely $-\kappa_\theta^{C}(p_S^{(\infty,H)} - p_C)$, is always non-negative.
Second, in Appendix~\ref{AppB} we show rigorously that $p_S^{(\infty,H)} + p_C \ge 1$; simulations are also presented in Appendix \ref{AppC}. It follows then that the second term, $\kappa_\phi^{C}(p_S^{(\infty,H)} - (1 - p_C))$, is also non-negative.
Since both contributions to Eq. (\ref{eq:QC}) are nonnegative,  $Q_C^{(\infty)}\geq 0$.
This constitutes one of the main results of this study: The alternating model cannot act as a quantum thermal refrigerator for any interaction energy or collision time.

Regarding the heat exchanged with the hot bath in the limit cycle, $Q_H^{(\infty)}$, this quantity can be positive, corresponding to heat flowing into the hot ancillas, or negative, corresponding to heat being extracted from them. 
The former scenario is associated with work being performed on the system through the switching of the interaction Hamiltonian, with this work subsequently dissipating into the baths. 
The scenario of negative heat exchange with the hot bath corresponds, for example, to a purely heat-transfer process from the hot bath to the cold bath, with no work contributions, which we describe next.

\begin{figure}[htbp]
    \centering
    \includegraphics[width=1\linewidth]{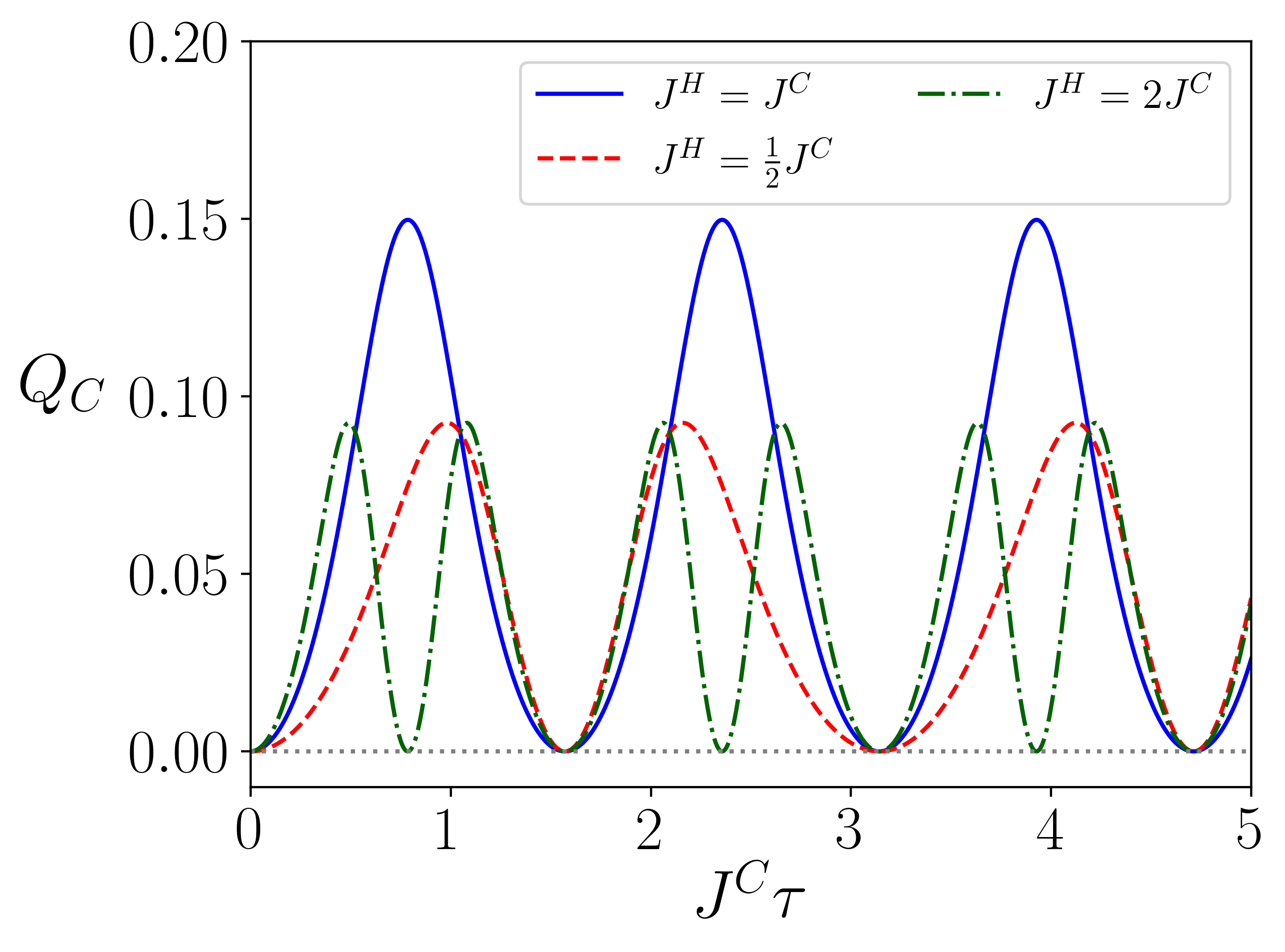}
\caption{Heat exchange at the limit cycle in the alternating-coupling model.  Shown is the heat exchanged in a single collision between the system and the cold ancilla at the limit cycle, as given by Eq.~(\ref{eq:QA}). 
We consider $J^{A}_{xx} = J^{A}_{yy} \equiv J$ with $A = H, C$,
and test cases with asymmetrical couplings at the two ends.
Other parameters are $\omega_{A} = \omega_{S} = 1$, $\tau = 0.5$, $\beta_{C} = 2$, and $\beta_{H} = 1$.} 
    \label{fig:Q}
\end{figure}

\begin{figure}[htbp]
    \centering
    \includegraphics[width=1\linewidth]{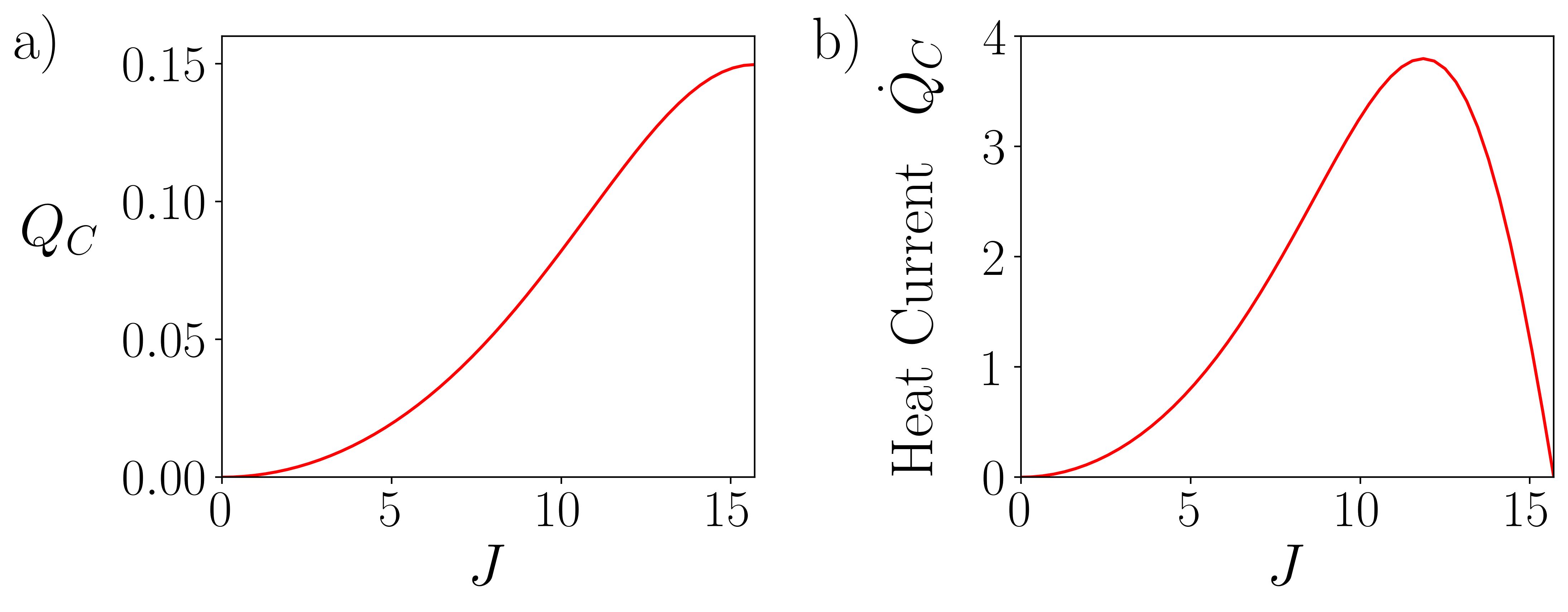}
\caption{Alternating-coupling model at limit cycle: (a) heat dissipated at the cold ancilla within each collision of duration $\tau$,   (b) heat current obtained as the $\tau$ derivative of panel (a).     We set $\tau=0.05$, $\omega_S=\omega_{A} = 1$, $\beta_C = 2$, and $\beta_H = 1$.
    }
    \label{fig:QC_EC}
\end{figure}


\subsubsection{Heat transfer without work}

We focus here on quantum heat transfer between two heat baths, from a hot reservoir to a cold one, in the absence of work done/extracted. At this small scale, the interaction energy between the system and its environment can play a central role. Understanding the behavior of the heat current in the strong-coupling regime has therefore been an active area of research.
Previous studies employing quantum master equations \cite{Segal05,Thoss10,Nicolin11,Segal14,Cao15,Cao17,Anto21}, Green’s function techniques \cite{Agarwalla17}, and numerically exact methods \cite{Thoss08,Nazim14,Kilgour19} have shown that the heat current is typically suppressed in the strong coupling limit. However, these studies were done assuming harmonic baths and interactions based on displacements of these harmonic modes.
Here, we investigate whether the present model with repeated-interaction (spin) baths and Heisenberg interaction can exhibit a suppression of the heat current in the strong-coupling regime.
Importantly, we can solve the model exactly analytically for arbitrary coupling strengths and collision times.

The model becomes purely heat-conducting (i.e., with no work contribution) when 
$[\hat H_S+\hat H_A,\hat H_I^{A}]=0$,
which holds here when $\omega_{S} = \omega_{A}$ and $J^{A}_{xx} = J^{A}_{yy}$, with $A = H, C$. In what follows, we use $\omega$ to denote the qubit splitting.
In this case, energy is transferred solely as heat flowing from hot ancillas to the cold one through the qubit system.  
The heat deposited into a cold ancilla at each collision is achieved from Eq.~(\ref{eq:Q}),
\begin{equation}
Q_{C}^{(\infty)} = \omega\,\kappa_{\theta}^{C}\bigl(p_{C} - p_{S}^{(\infty, H)}\bigr)\:.
\end{equation}
This expression has a clear physical interpretation: the exchanged heat is proportional to the population imbalance between the cold ancilla and the system population immediately after the preceding collision (with a hot ancilla). 
Moreover, heat exchange scales linearly with the coupling coefficient $\kappa_{\theta}^{C}$ and increases with the quantum of energy $\omega$.

Using the limit cycle populations, Eq. (\ref{eq:pop}), this expression simplifies to
\bea
    Q_{C}^{(\infty)} =\frac{\omega \kappa_\theta^C \kappa_\theta^H }{\kappa_\theta^C \kappa_\theta^H-\kappa_\theta^C-\kappa_\theta^H}(p_H-p_C),
\label{eq:QA}
\eea
where $\kappa_{\theta}^A =\sin ^2\left(2 J^A \tau \right)$.

If we further consider the symmetric coupling case, $J^C = J^H\equiv J$, we get
\begin{equation}
\label{eq:simp-heat-C-alternating}
    Q_{C}^{(\infty)} =\frac{\omega \sin ^2\left(2 J  \tau \right)  }{2-\sin ^2\left(2 J  \tau \right)  }(p_C-p_H).
\end{equation}
The heat exchanged at each collision reaches a minimum when $2J\tau = n\pi$, with $n \in \mathbb{Z}$, 
and is maximized at intermediate values, namely when $4J\tau = n\pi$. 
The behavior is illustrated in Fig.~\ref{fig:Q}. 
More complex, yet still periodic, patterns emerge when $J^{C} \neq J^{H}$, as we show in the same Figure.
Such periodic trends are a generic feature of the repeated interaction model since the effective coupling strength is governed by the periodic function  $0 \leq \sin^{2}(2J\tau) \leq 1$, see Eq.~(\ref{eq:kappatp}). 
As this effective coupling function is tuned from $0$ to $1$, the heat exchanged per collision increases monotonically as we show in Fig. \ref{fig:QC_EC}(a).

We next derive an expression for the \emph{heat current} by differentiating the heat exchanged per collision, $Q_{C}^{(\infty)}$, with respect to the collision time $\tau$. Assuming for simplicity equal couplings to the H and C baths, we obtain
\bea
\dot Q_C^{(\infty)} = 4J\omega\frac{\sin\left(4J\tau\right)}{[2- \sin^2\left(2J\tau\right)]^2} (p_C-p_H).
\label{eq:dQA}
\eea
We can express Eq. (\ref{eq:dQA}) in terms of an effective coupling $x=\sin^2 \left(2J\tau\right)$, resulting in 
\begin{equation}
    \dot Q_C^{(\infty)}= 8J\omega\frac{\sqrt{x(1-x)}}{(2-x)^2} (p_C-p_H).
    \label{eq:dQA-x}
\end{equation}
This function exhibits a nonmonotonic {\it turnover} behavior as a function of $x$ and $J$: It initially increases with $x$, reaches a maximum at intermediate coupling, and is subsequently suppressed to zero in the strong-coupling regime; see Fig.~\ref{fig:QC_EC}(b).

This behavior is consistent with previous studies reporting a turnover behavior at strong coupling, particularly for systems coupled to bosonic baths and commonly analyzed within quantum master equation frameworks \cite{Segal05,Nicolin11,Anto21,Segal14,Cao15,Cao17,Thoss10}. 
However, in contrast to conventional spin-boson models, a qubit in the repeated-interaction setting does not exhibit thermal rectification \cite{Segal05,Segal14}, as is evident from Eq.~(\ref{eq:QA}).

Equation~(\ref{eq:dQA}), or equivalently, Eq.~(\ref{eq:dQA-x}) constitutes another key result of our study. It is an exact expression for the heat current in the RI model (with all interaction coefficients identical), valid for arbitrary coupling strength and collision time. It demonstrates that the thermal conductance of the junction displays pronounced nonlinear and non monotonic behavior at strong coupling, culminating in a complete suppression of the heat current in the ultrastrong-coupling limit.

In the weak coupling limit, $J \tau \ll 1$, we expand Eq. (\ref{eq:dQA}) in a Taylor series to the second order. 
The heat deposited to the cold ancilla is given by
\begin{equation}
    Q_{C}^{(\infty)} = 2 \omega (p_C-p_H)(J \tau)^2.
\end{equation}
Hence, we find the heat current to be
\begin{equation}
    \dot{Q}_C^{(\infty)} = 2 \omega (p_C-p_H)J^2 \tau.
\end{equation}
Note that in this limit, $\tau \to 0^{+}$ and $J\tau \to 0$, while $J^{2}\tau \to \Gamma$, a constant. 
As expected on intuitive grounds, in the weak-coupling regime, the heat flowing into the cold bath is proportional to the ancilla's frequency (equal to the system's), the difference in ground-state populations between a cold and a hot ancilla, reflecting the temperature bias, and the rate constant $\Gamma$, which to lowest order in coupling depends on both the collision duration and the coupling strength.

\begin{figure}[h]
    \centering
    \includegraphics[width=1\linewidth]{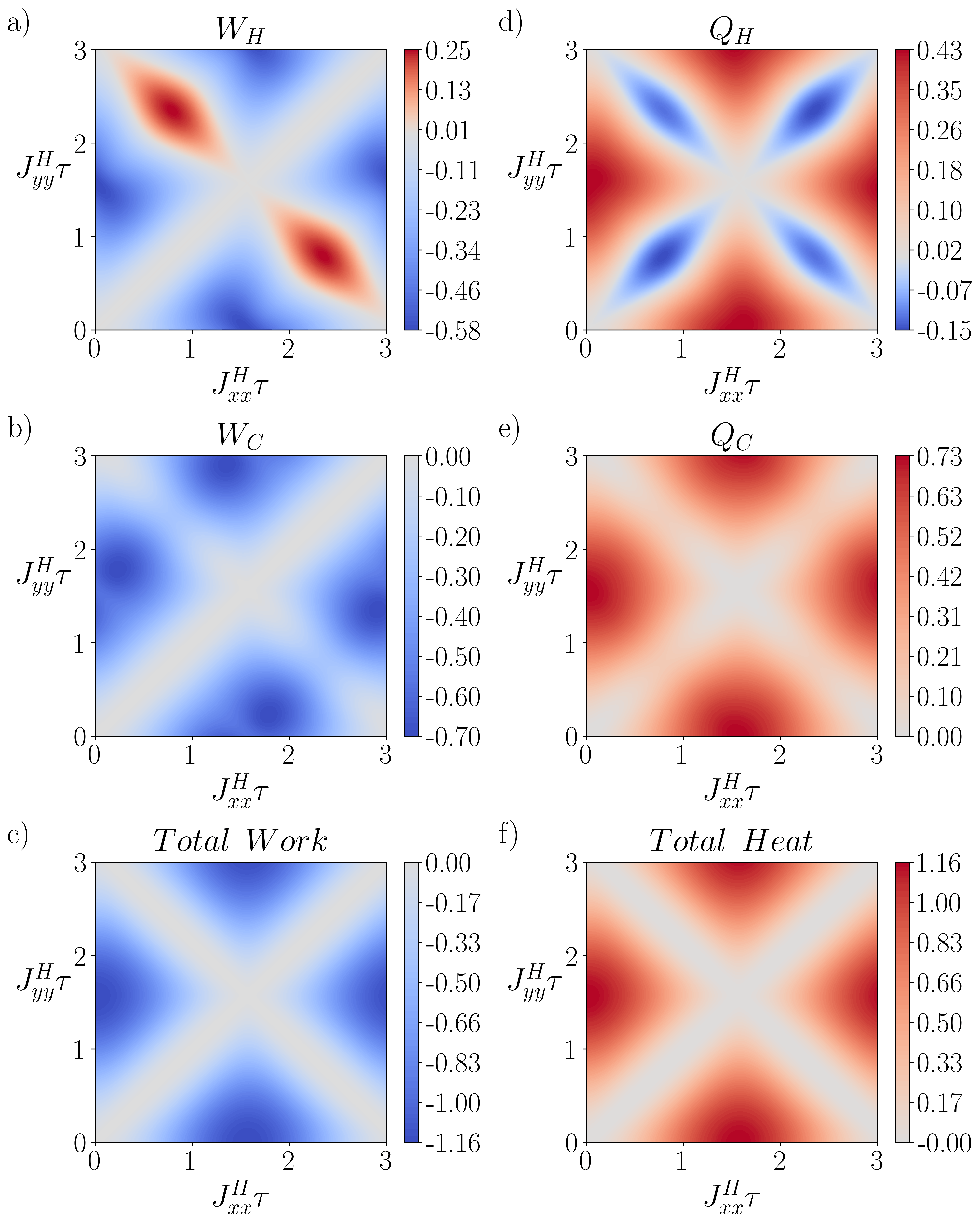}
    \caption{Alternating-coupling model. (a)-(c) Work and (d)-(f) heat exchange at the hot (top panels) and cold (middle panels) ancillas at limit cycle  for different values of the interaction strengths $J_{xx}$,\:$J_{yy}$.  
    The bottom panels present the  (c) work and (f) heat, when summed up over both hot and cold strokes.
    We set the coupling parameters to be hot-cold symmetric, $J_{xx}^{C}=J_{xx}^{H}$ and $J_{yy}^{C}=J_{yy}^{H}$. Other parameters  are $\omega_{A} = \omega_{S}=1$, $\tau=0.5$,  $\beta_{C}=2$ and $\beta_{H}=1$.    }
    \label{fig:mapA1}
\end{figure}

\begin{figure}
    \centering
    \includegraphics[width=1\linewidth]{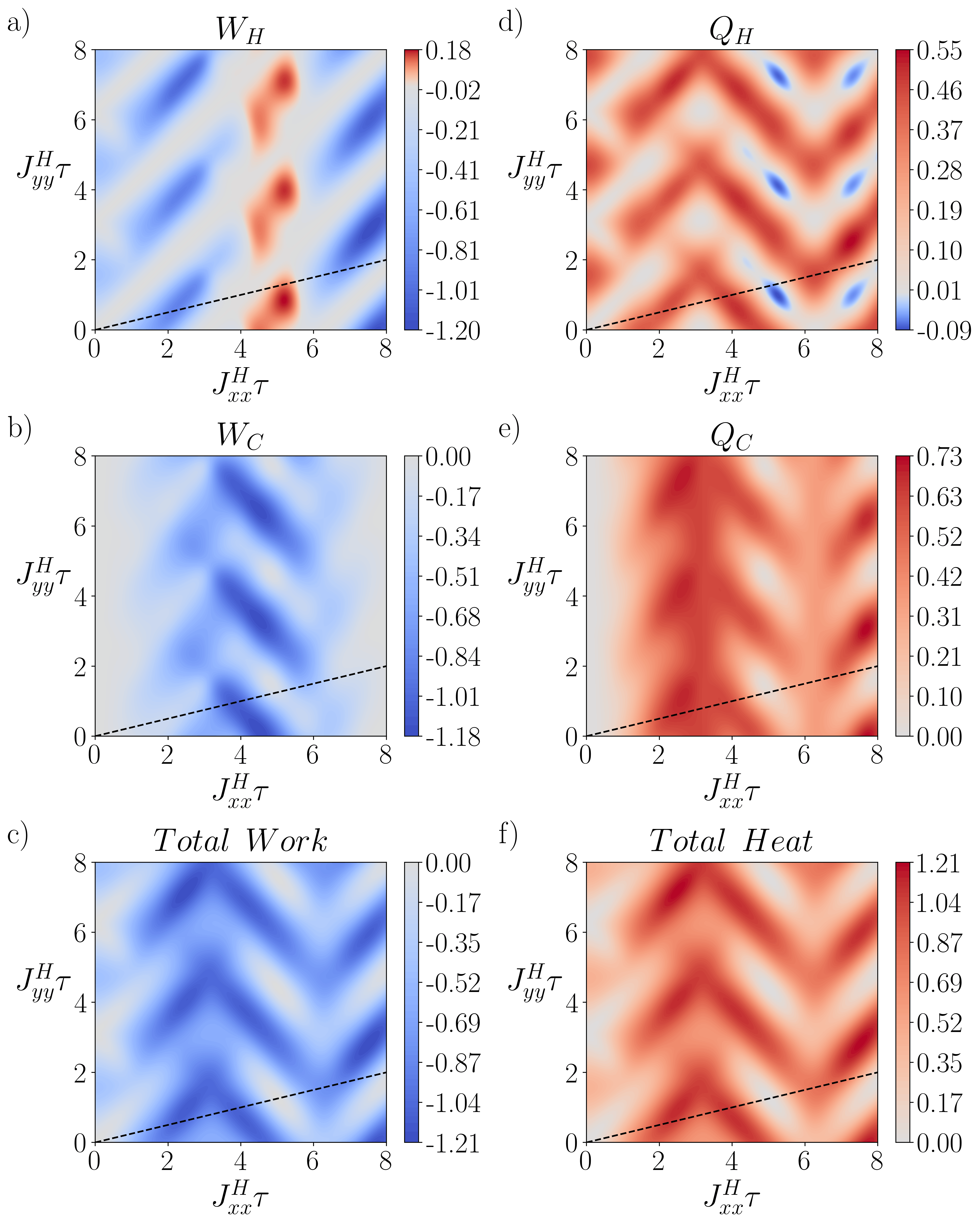}
    \caption{Alternating-coupling model. (a)-(c) Work and (d)-(f) heat exchange to the hot (top panels) and cold (middle panels) ancillas at limit cycle  for different values of the interaction strengths $J_{xx}^{H},\:J_{yy}^{H}$. 
      The bottom panels present the (c) work and (f) heat, when summed up over both hot and cold strokes.
    We use asymmetric couplings, $J_{xx}^{C}=\frac{1}{2}J_{xx}^{H}$ and $J_{yy}^{C}=\frac{1}{8}J_{xx}^{H}$. Other parameters are $\omega_{A}=\omega_{S}=1$, $\tau=0.5$, $\beta_{C}=2$ and $\beta_{H}=1$.
    The dashed line follows the $J_{yy}^{H}=\frac{1}{4}J_{xx}^{H}$ cut; it corresponds to the scenario considered in Fig. \ref{fig:W}.}
    \label{fig:mapA2}
\end{figure}

\begin{figure}[htbp]
    \centering
\includegraphics[width=1\linewidth]{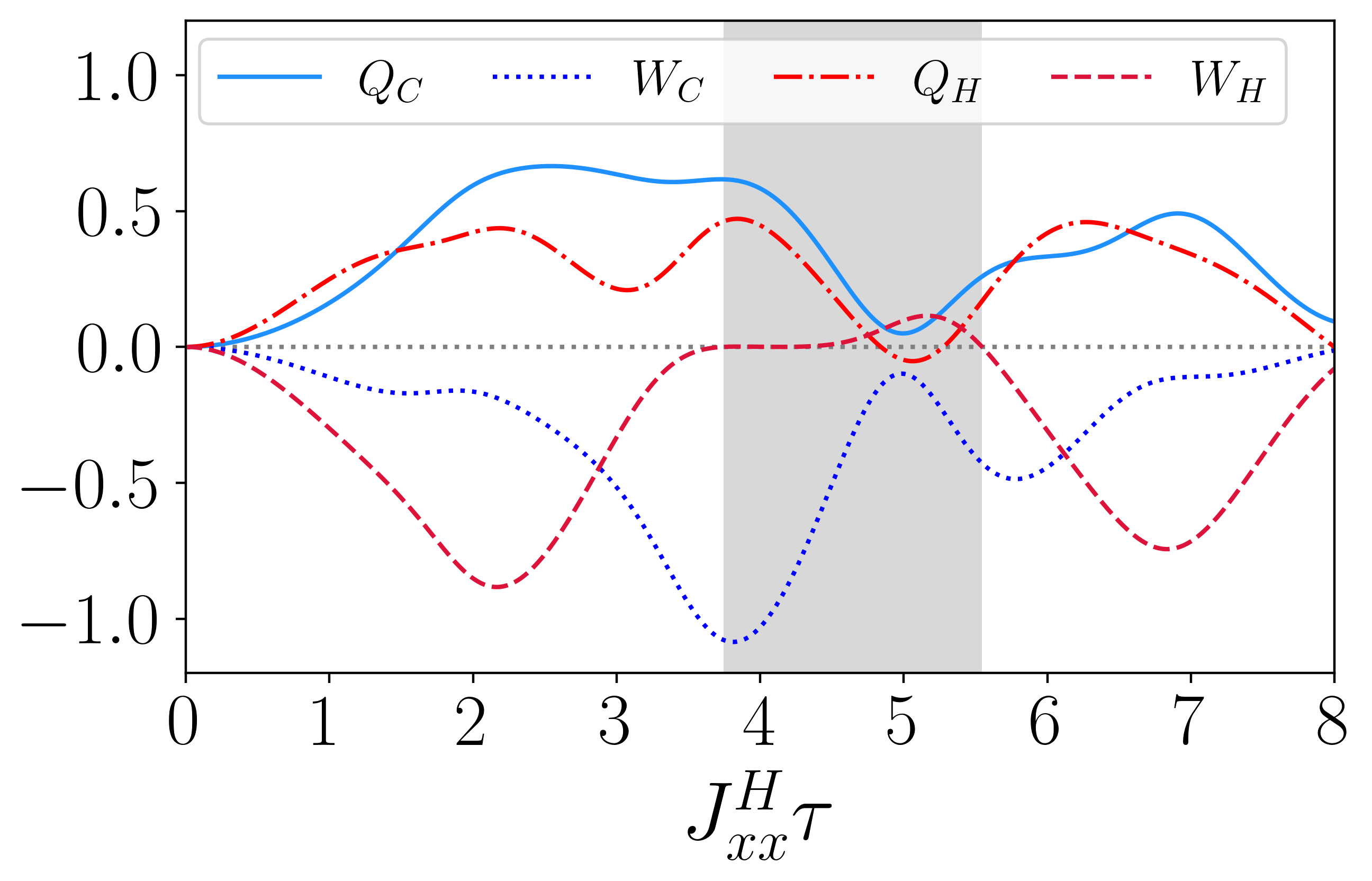}
    \caption{Alternating-coupling model. We present thermodynamical functions in the limit cycle as a function of $J_{xx}^{H}\tau$. Parameters are $J_{yy}^{H} =\frac{1}{4}J_{xx}^{H}$, $J_{xx}^{C}=\frac{1}{2}J_{xx}^{H}$, $J_{yy}^{C}=\frac{1}{8}J_{xx}^{H}$, $\tau=0.5$, $\beta_{C}=2$ and $\beta_{H}=1$. The grey region highlights areas where $W_{H}>0$.}
    \label{fig:W}
\end{figure}

\subsubsection{Work generation at the hot contact, $W_H>0$}

We are interested in exploring modes of operation of this nonequilibrium-driven quantum system. 
To this end, we scan the system’s performance over a broad range of parameters, varying the four interaction parameters.
In Figs. \ref{fig:mapA1}-\ref{fig:mapA2} we present comprehensive displays of the performance of the system across different choices of coupling parameters. Fig. \ref{fig:mapA1} enforces the symmetric coupling to the two baths, i.e. $J_{xx}^C=J_{xx}^H$ and $J_{yy}^C=J_{yy}^H$, while in Fig. \ref{fig:mapA2} we relax this symmetry.
In both cases, $Q_H^{(\infty)}$ can be negative in some regions, but, as expected, $Q_C^{(\infty)}$ is positive throughout without refrigeration function. Furthermore, work on the hot side is shown to be positive in some parameter regimes $W_H^{(\infty)}>0 $, which means that the system performs work on the hot contact. Note that according to Eq. (\ref{eq:W}), if $\omega_S=\omega_A$, which is our typical setting, then  $W_C^{(\infty)}$ is always negative. 
Regarding the overall work generated by the system, $W_H^{(\infty)}+W_C^{(\infty)}$, extensive simulations using optimization approaches have shown that it is always negative. That is, the system consumes interaction energy, which is then released as heat to the baths' ancillas, and no engine operation is realized.

To further display trends, we present in Fig.~\ref{fig:W} a cut along Fig. \ref{fig:mapA2}, highlighting the regime in which $W_{H}^{(\infty)} > 0$.  This work is generated as an interaction energy at the hot interface at the expense of heat extracted from the hot bath $Q_{H}^{(\infty)} < 0$ and work drawn from the interaction energy with the cold ancillas, $W_{C}^{(\infty)} < 0$. Once again, we note that, on the basis of simulations, the total work is always negative.

\subsection{Summary of performance: Alternating coupling thermal device}
\label{sec:IID}

In this section, we presented an exact analysis of a qubit-based thermal machine operating under an alternating RI protocol, focusing on its limit-cycle performance. 
We obtained closed-form expressions for the system’s limit-cycle state, energy flow, and thermodynamic performance, valid for arbitrary interaction strengths and collision durations.

{\it I. State of the system.}
Our first main result is the exact characterization of the limit cycle. We derived explicit expressions for the system ground-state population immediately after collisions with either the hot or cold ancillas, Eqs.~(\ref{eq:pop}); coherences decay to zero. 

{\it II. No-go theorem for refrigeration.}
We established rigorous bounds on populations of the system, demonstrating that the system can never become colder than the cold bath, see Eq. (\ref{eq:boundpop}), irrespective of coupling strengths or collision times. These bounds lead directly to a {\it no-go theorem for refrigeration}. We proved that the heat exchanged with the cold bath in the limit cycle is always nonnegative, $Q_C^{(\infty)} \ge 0$, ruling out the possibility of extracting heat from the cold reservoir. Consequently, the alternating RI model cannot operate as a quantum thermal refrigerator under any choice of parameters. This result holds beyond the weak-coupling or stroboscopic (fast collision) regime.

{\it III. Heat conduction.}
We analyzed the behavior of the system as a {\it heat conductor} in the absence of work. The heat exchanged per collision exhibits a characteristic oscillatory dependence on the coupling strength and collision time, reflecting the unitary nature of individual collisions. Notably, the heat current displays a pronounced turnover behavior: it increases at weak coupling, reaches a maximum at intermediate coupling, and is fully suppressed in the ultrastrong-coupling regime.

{\it IV. Work contributions.}
We investigated the work contributions associated with switching the system-ancilla interactions. We identified parameter regimes in which positive work is generated locally at the hot contact, but proved that the work is always nonpositive at the cold side. An extensive numerical exploration confirmed that the net work over a cycle is always negative, implying that the device cannot function as a heat engine.

Altogether, results of this section show that even when allowing for strong coupling and long collision time, thus deviating from the standard Lindbladian dynamics, the model cannot realize refrigeration nor engine operation. These findings should motivate the exploration of enriched models, non-Markovian ancilla correlations, or multilevel working fluids.

\section{Simultaneous Coupling Thermal Machine}
\label{sec:Simult}

\begin{figure}[h]
    \centering
    \includegraphics[width=0.9\linewidth]{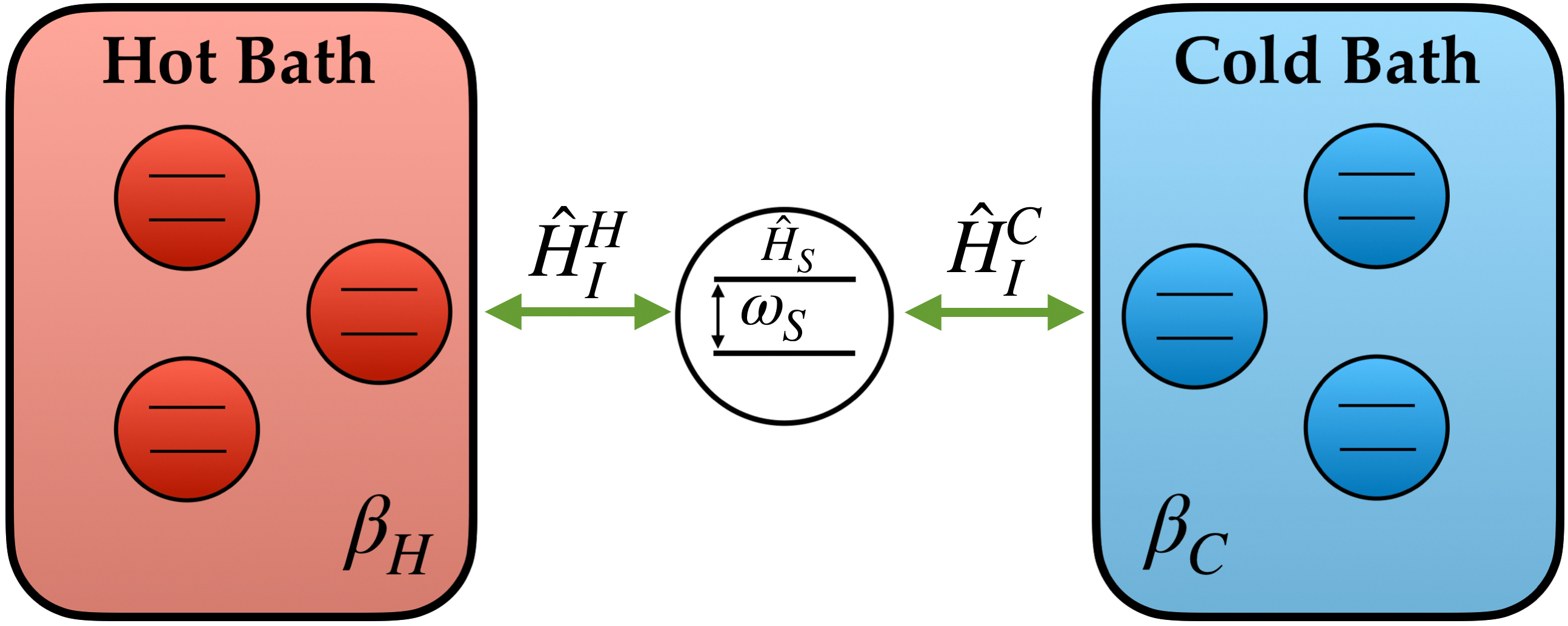}
    \caption{Simultaneous-coupling model for a quantum thermal device. A two level system ($\hat H_S$) collides with both the cold and the hot ancillas, simultaneously.} 
    \label{fig:SketchS}
\end{figure}

We now consider a second model of a quantum thermal machine within the RI framework. In contrast to the model introduced in Sec. \ref{sec:Alter}, the qubit system is allowed here to interact simultaneously with both baths, represented by hot and cold ancillas. A schematic illustration of this setup is shown in Fig. \ref{fig:SketchS}.

We analyze the model numerically for general coupling energy and collision times, beyond the ``short time" Lindblad limit \cite{RamonEscandel2025}.
Analytical results are presented here in the limit of short collision time $J\tau\ll1$, without placing restrictions on the qubit and the ancilla energy gaps, $\omega$. If we further impose a short time limit with respect to the free system and the ancilla's evolution, $\omega\tau\ll1$, one can prove that the simultaneous coupling thermal machine and the alternating machine operate identically.

A collision now includes a three-body interaction Hamiltonian, given by
\begin{equation}
\begin{split}
    \hat{H}_{I} &= J_{xx}^{H}\hat{\sigma}_{x}^{S}\otimes\hat{\sigma}_{x}^{H}\otimes \unit^{C} + J_{yy}^{H}\hat{\sigma}_{y}^{S}\otimes\hat{\sigma}_{y}^{H}\otimes \unit^{C} \\
    &+ J_{xx}^{C}\hat{\sigma}_{x}^{S}\otimes \unit^{H}\otimes\hat{\sigma}_{x}^{C}+ J_{yy}^{C}\hat{\sigma}_{y}^{S}\otimes \unit^{H}\otimes\hat{\sigma}_{y}^{C}.
\end{split}
\end{equation}
The total RI Hamiltonian is
\begin{equation}
    \begin{split}
        \hat{H}_{tot} &= \hat{H}_{S}\otimes\unit^{H}\otimes\unit^{C}+\unit^{S}\otimes\hat{H}_{H}\otimes\unit^{C}\\
        &+\unit^{S}\otimes\unit^{H}\otimes\hat{H}_{C} + \hat{H}_{I}\:,
    \end{split}
\end{equation}
providing the collision unitary
\begin{equation}
    \hat{U}(\tau) = e^{- i \hat{H}_{tot} \tau}\:.
    \label{eq:UTS}
\end{equation}
In the short collision time limit,   
$\hat U(\tau)\approx\hat U_H(\tau)\hat U_C(\tau) \hat U_S^{\dagger}(\tau)$,
with $\hat U_A$ defined for the alternating model, Eq. (\ref{eq:UA}).
The error in this trotterization scales as a product of the characteristic interaction, the frequency gap, and the time interval squared, namely $O(J\omega\tau^2)$. Here, $\hat U_S(\tau)$ describes the free evolution of the system's qubit. This free evolution term only adds a non-influential phase factor to coherences; it does not change neither the population dynamics nor the thermodynamical function of the machine. Thus, in this short time limit, the two machines, whether operated in an alternating manner or with the baths coupled simultaneously, behave identically.

\subsection{Population evolution: Perturbative solution}
Returning to the general time evolution, Eq. (\ref{eq:UTS}), after each collision, the state of the system evolves according to
\begin{equation}
    \rho_S^{(n+1)} = \text{Tr}_{H,C}\Big[\hat{U}(\tau) (\rho_S^{(n)} \otimes \rho_H  \otimes \rho_C )\hat{U}^{\dagger}(\tau)\Big]\:\:.
\label{eq:Sim_model_map}
\end{equation}
In this scenario, obtaining an exact analytical solution for the general dynamics becomes cumbersome. However, in the limit of short collision times, $\tau \ll J^{-1}$, we derive an approximate analytical solution without imposing restrictions on the system or ancilla frequencies $\omega_{\alpha}$. This is achieved by expanding the collision unitary $\hat{U}(\tau)$ as a Dyson series up to the second order in $\tau$. For the system ground state population, we get 
\begin{equation}
    p_{S}^{(n+1)} = \eta\left(p_{S}^{(n)}-p_{S}^{(\infty)}\right) + p_{S}^{(\infty)},
    \label{eq:pop_ansatz}   
\end{equation}
with the coefficient
\bea
\label{eq:eta_short_tau}
    \eta = 1-2\left[\left(J_{xx}^{H}\right)^{2}+\left(J_{yy}^{H}\right)^{2}+\left(J_{xx}^{C}\right)^{2}+\left(J_{yy}^{C}\right)^{2}\right]\tau^{2} +O(\tau^{3}),
    \nonumber\\
\eea
and the fixed point
 \bea
   && p_{S}^{(\infty)} = 
   \nonumber\\
   &&\frac{\left(J_{xx}^{C}-J_{yy}^{C}\right)^{2}+\left(J_{xx}^{H}-J_{yy}^{H}\right)^{2} +4J_{xx}^{C}J_{yy}^{C}p_{C}+4J_{xx}^{H}J_{yy}^{H}p_{H}}{2\left[\left(J_{xx}^{H}\right)^{2}+\left(J_{yy}^{H}\right)^{2}+\left(J_{xx}^{C}\right)^{2}+\left(J_{yy}^{C}\right)^{2}\right]},
    \nonumber\\
    \label{eq:Dyson-sim-ss}
\eea
as steady state population at long times.
As shown in Eq. (\ref{eq:popAdt}), the alternating and the continuous models reach the same limit cycle solution for the population in the short $\tau$ limit. 

A nonperturbative analytical solution of the dynamics described by Eq. (\ref{eq:Sim_model_map}) can be obtained, independently of the value of $\tau$, in the special case where $J_{xx}^{A} = J_{yy}^{A} \equiv J$ with $A = H, C$. In this case, the population of the system after a single collision evolves as
\begin{equation}
\begin{split}
    p_S^{(n+1)} &= \frac{1}{4} \bigg[-\cos \left(4 \sqrt{2} J \tau \right) (p_C+p_H-2
   p_S^{(n)})\\&+p_C+p_H+2 p_S^{(n)}\bigg],
\end{split}
\end{equation}
and the steady state is
\begin{equation}
    p_S^{(\infty)} = \frac{p_C + p_H}{2}.
\end{equation}
Namely, the system population reaches an intermediate value between the two baths' temperatures.

\begin{figure*}[ht]
\centering
\includegraphics[width=1\linewidth]{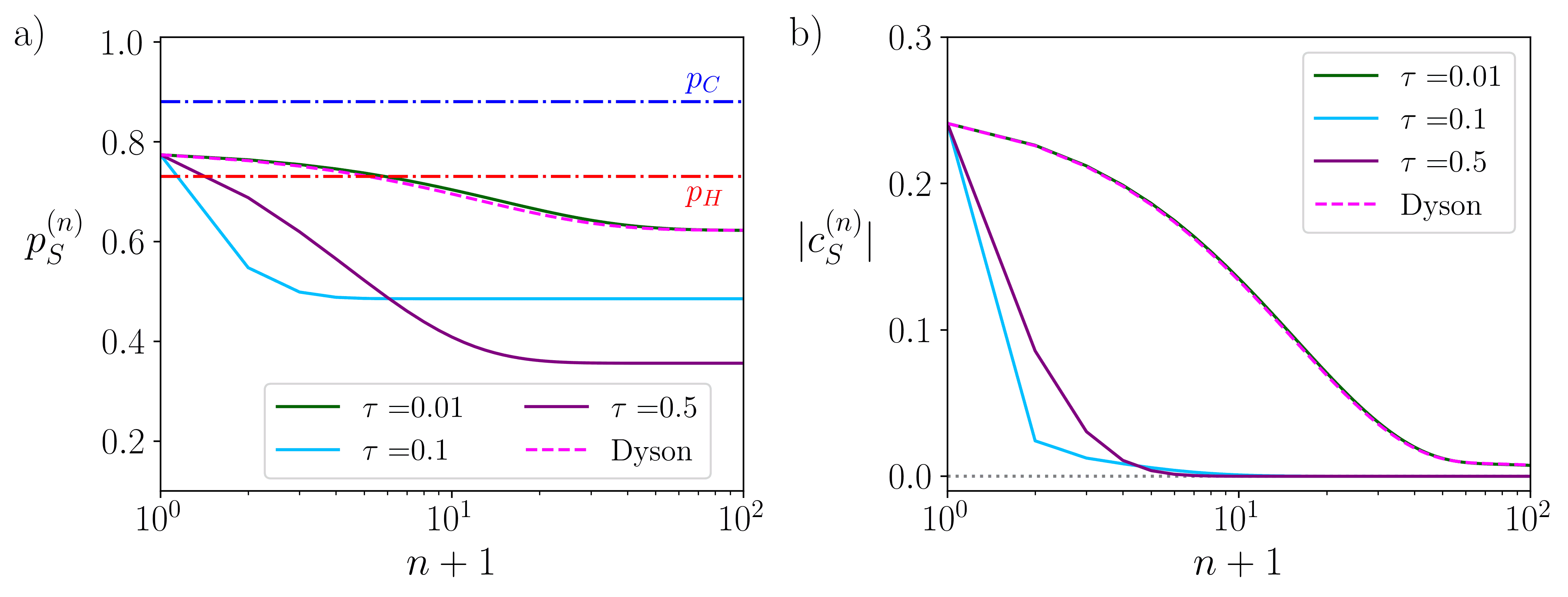}
\caption{System dynamics in the simultaneous-coupling model. 
(a) Ground state population and (b) coherence evolution both plotted as a function of the number of collisions. 
We present numerical simulations for $\tau=0.01, 0.1, 0.5$ with analytical results for the short collision time limit (dashed magenta) based on Eqs. (\ref{eq:Dyson-sim-ss}) and (\ref{eq:Dyson-psi}). Parameters are $J^C_{xx} = 2$, $J^C_{yy} = 8$, $J^H_{xx} = 4$, $J^H_{yy} = 16$, $\omega_{S}=\omega_{A}$, $\beta_C = 2$, and $ \beta_H = 1$.}
\label{fig:SimultaneousModel_dynamics}
\end{figure*}

In Fig. \ref{fig:SimultaneousModel_dynamics}(a), we display the population relaxation dynamics for three different values of the collision time $\tau$. Although Eqs. (\ref{eq:eta_short_tau}) and (\ref{eq:Dyson-sim-ss}) exhibit no explicit dependence on $\tau$, they are derived, and remain valid, only in the short-collision limit, $J\tau\ll1$, referred to as the ``Dyson" solution, as we demonstrate in Fig. \ref{fig:SimultaneousModel_dynamics}(a).
Moreover, we note that the system steady-state population reached at long times, $p_{S}^{(\infty)}$, can be lower than the ground-state population of the hot ancilla, $p_{H}$. This implies that the system can become hotter than both baths to which it is coupled.

To gain some insight about the conditions under which this happens, we focus on the inequality $p_{S}^{(\infty)}<p_{H}$. This is equivalent [see Eq. (\ref{eq:Dyson-sim-ss})] to
\begin{equation}
\begin{split}
    &\left(1-2p_{H}\right)\left[(J_{xx}^{H}-J_{yy}^{H})^{2}+(J_{xx}^{C}-J_{yy}^{C})^{2}\right]\\&+4J_{xx}^{C}J_{yy}^{C}(p_{C}-p_{H})<0\:\:.
\end{split}
\end{equation}
Since the second line is always positive and $1-2p_H<0$,
it is clear that a necessary condition for the system ``overheating" is that $J_{xx}^A\neq J_{yy}^A$, which is linked to work generation due to the on/off process of the interaction Hamiltonian.

We can further derive an equation of motion for the system ground state population in the simultaneous model, assuming short collision times. 
Based on Eq. (\ref{eq:pop_ansatz}), we get
\bea
&&\frac{p_{S}^{(n+1)}-p_{S}^{(n)}}{\tau} = 
\nonumber\\&&-2\tau\left[\left(J_{xx}^{H}\right)^{2}+\left(J_{yy}^{H}\right)^{2}+\left(J_{xx}^{C}\right)^{2}+\left(J_{yy}^{C}\right)^{2}\right]p_{S}^{(n)}
\nonumber\\
&&+2\tau\left[\left(J_{xx}^{H}\right)^{2}+\left(J_{yy}^{H}\right)^{2}+\left(J_{xx}^{C}\right)^{2}+\left(J_{yy}^{C}\right)^{2}\right]p_{S}^{(\infty)},
\label{eq: pop_pre_EOM}
\eea
which can be written in the limit $\tau\rightarrow0^{+}$ limit as
\bea
\dot{p}_{S}(t) &=& -2\left(\Gamma_{xx}^{C}+\Gamma_{yy}^{C}+\Gamma_{xx}^{H}+\Gamma_{yy}^{H}\right)p_{S}(t)
\nonumber\\
&+& 2\left(\Gamma_{xx}^{C}+\Gamma_{yy}^{C}+\Gamma_{xx}^{H}+\Gamma_{yy}^{H}\right)p_{S}^{(\infty)},
\label{eq:EOM_populations}
\eea
with the rates defined as $\Gamma_{xx}^A=(J_{xx}^A)^2\tau$
and $\Gamma_{yy}^A=(J_{yy}^A)^2\tau$.
\subsection{Decoherence dynamics: Perturbative solution}
For the coherences, writing $c_{S}^{(n)}$ in polar form as $c_{S}^{(n)} = \lvert c_{S}^{(n)} \rvert e^{i\chi}$ and noting that $c_{S}^{(n+1)} = \psi \lvert c_{S}^{(n)} \rvert$, the Dyson expansion yields
\begin{equation}
\begin{split}
    \psi &= \bigg[1+i\omega\tau\\&-\left.\left(\frac{\omega^{2}}{2}+\left(J_{xx}^{C}\right)^2+\left(J_{yy}^{C}\right)^2+\left(J_{xx}^{H}\right)^2+\left(J_{yy}^{H}\right)^2\right)\tau^{2}\right]e^{i\chi} \\&+\bigg[\left(J_{xx}^{C}\right)^2-\left(J_{yy}^{C}\right)^2+\left(J_{xx}^{H}\right)^2-\left(J_{yy}^{H}\right)^2\bigg]\tau^2 e^{-i\chi}+O(\tau^{3}).
\end{split}
\label{eq:Dyson-psi}
\end{equation}
We represent the decoherence dynamics in Fig. \ref{fig:SimultaneousModel_dynamics}(b). As previously observed for populations, the Dyson approximation accurately replicates the dynamics only when $J\tau\ll1$.

We also derive an equation of motion for coherences. 
In the limit $\tau\rightarrow{0}^{+}$, we get
\bea
    \dot{c}_{S}(t) &=& \left(i\omega-\Omega-\Gamma_{xx}^{C}-\Gamma_{yy}^{C}-\Gamma_{xx}^{H}-\Gamma_{yy}^{H}\right)c_{S}(t)
    \nonumber\\
    &+&\left(\Gamma_{xx}^{C}-\Gamma_{yy}^{C}+\Gamma_{xx}^{H}-\Gamma_{yy}^{H}\right)(c_{S}(t))^{*},
\eea
where we define $\Omega:=\frac{\omega\tau}{2}$ such that $\Omega$ is constant when $\tau\rightarrow0{^+}$.

\subsection{Thermodynamical analysis in the limit cycle}
\subsubsection{Heat exchange and quantum refrigeration}

We define the heat exchanged in a single collision between the system and the ancilla $A$ as
\begin{equation}
    Q_{A}^{(n+1)} = \text{Tr}\left[\left(\hat{U}^{\dagger}(\tau)\hat{H}_{A}\hat{U}(\tau)-\hat{H}_{A}\right)\rho_{S}^{(n)}\otimes\rho_{H}\otimes\rho_{C}\right],
\label{eq:Sim_Model_Heat_Def}
\end{equation}
with $A=H,C$.
%
 
\begin{figure}[ht]
\centering
\includegraphics[width=1\linewidth]{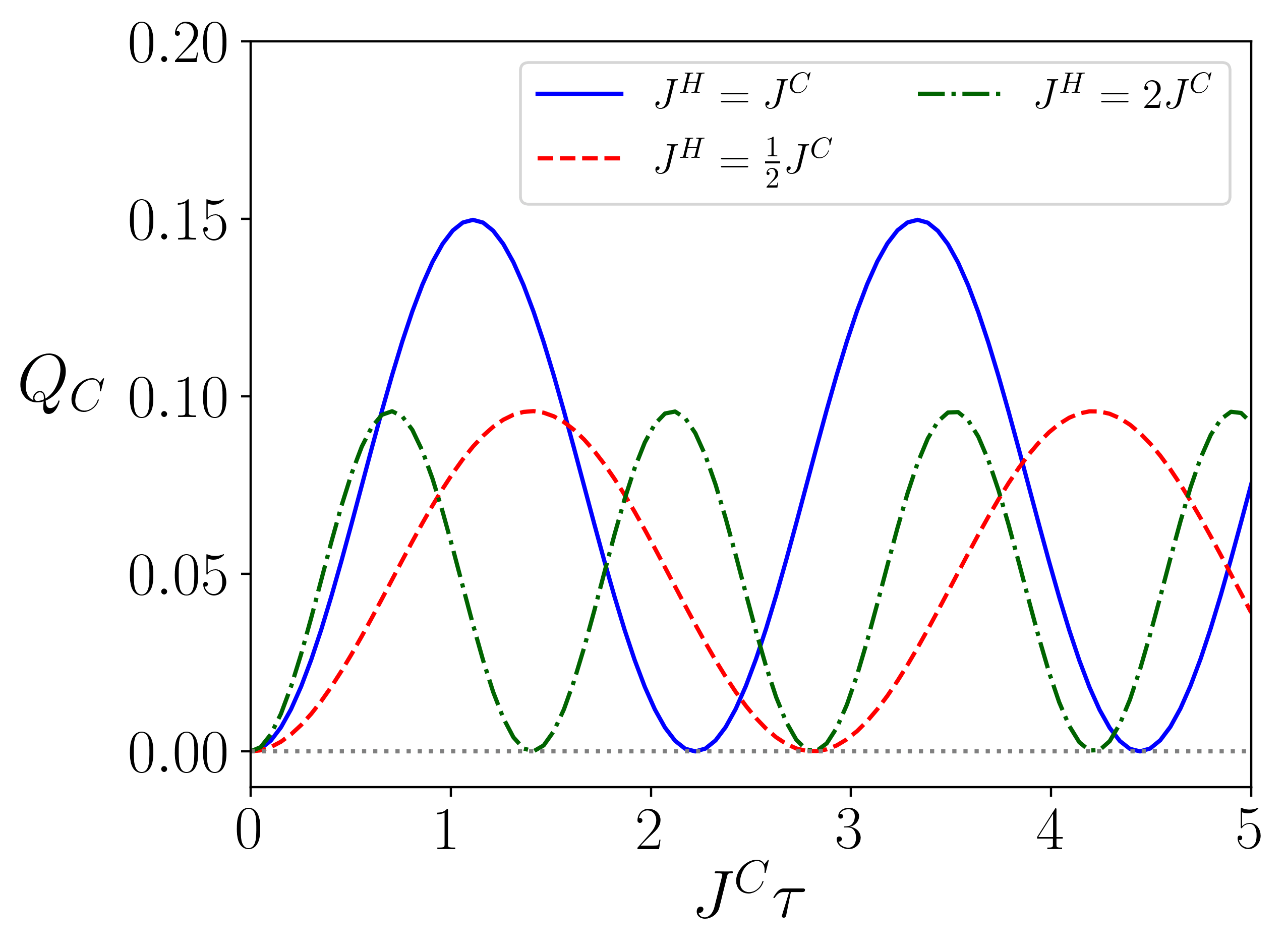}
\caption{Simultaneous-coupling model: 
Heat exchange between the cold bath and the system at steady state for a purely heat conducting model. 
We present results for both the symmetric, $J^C=J^H$, and the asymmetric, $J^C\neq J^H$, cases.
Other parameters are $\tau = 0.5$, $\omega_A =\omega_S = 1$, $\beta_H = 1$, and $\beta_C = 2$. 
}
   \label{fig:Heat1S}
\end{figure}

\begin{figure}[htbp]
 \centering
\includegraphics[width=1\linewidth]{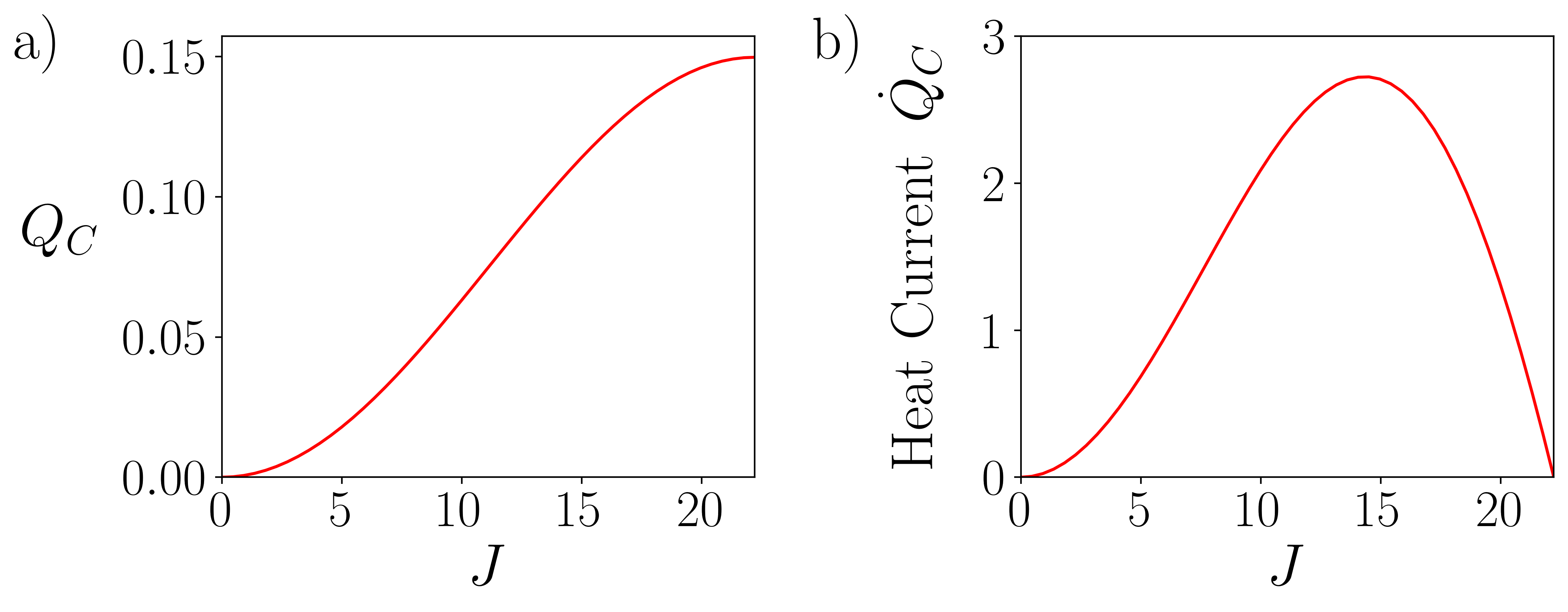}
\caption{Simultaneous-coupling model: heat exchange between the system and the cold ancilla at limit cycle (a) and the corresponding heat current (b). We set $\tau=0.05$, $\omega_{A} = 1$, and $\beta_C = 2 \;(\beta_H = 1)$ as inverse temperature of the cold (hot) bath.}
\label{fig:Heat2S}
\end{figure}


\begin{figure}
\centering
\includegraphics[width=1\linewidth]{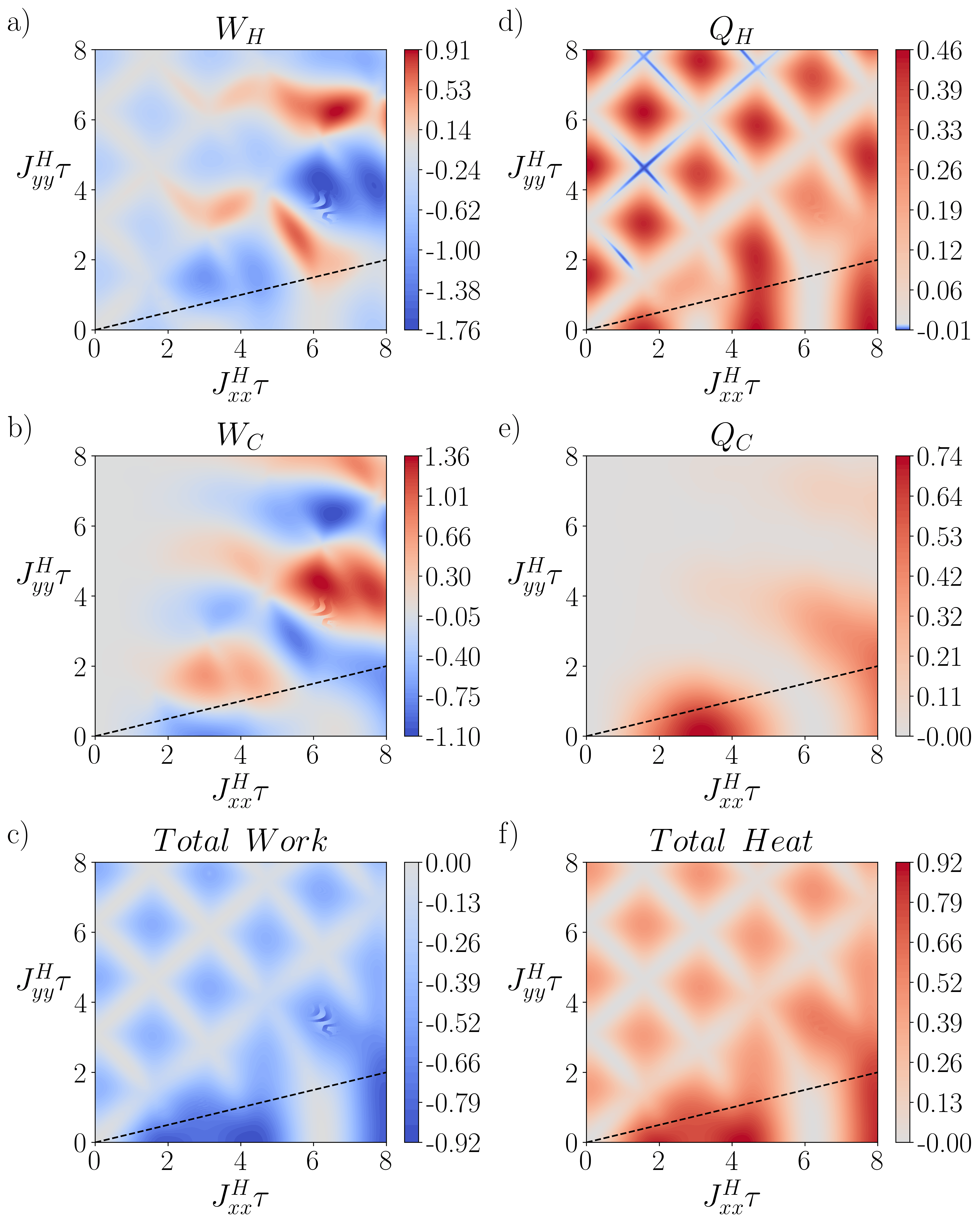}
\caption{Simultaneous-coupling model: (a)-(c) Work and (d)-(f) heat exchange to the hot (top panels) and cold (middle panels) ancillas at limit cycle for different values of the interaction strengths $J_{xx}^{H},\:J_{yy}^{H}$.
The bottom panels present the total (c) work and (f) heat from the two contacts.
We use asymmetric couplings, $J_{xx}^{C}=\frac{1}{2}J_{xx}^{H}$ and $J_{yy}^{C}=\frac{1}{8}J_{xx}^{H}$. Other parameters are $\omega_{A}=\omega_{S}=1$, $\tau=0.5$, $\beta_{C}=2$, and $\beta_{H}=1$. The dashed line follows the $J_{yy}^{H}=\frac{1}{4}J_{xx}^{H}$ cut; it corresponds to the scenario considered in Fig. \ref{fig:cutS}.}
\label{fig:mapS1}
\end{figure}

\begin{figure}[h]
    \centering
\includegraphics[width=1\linewidth]{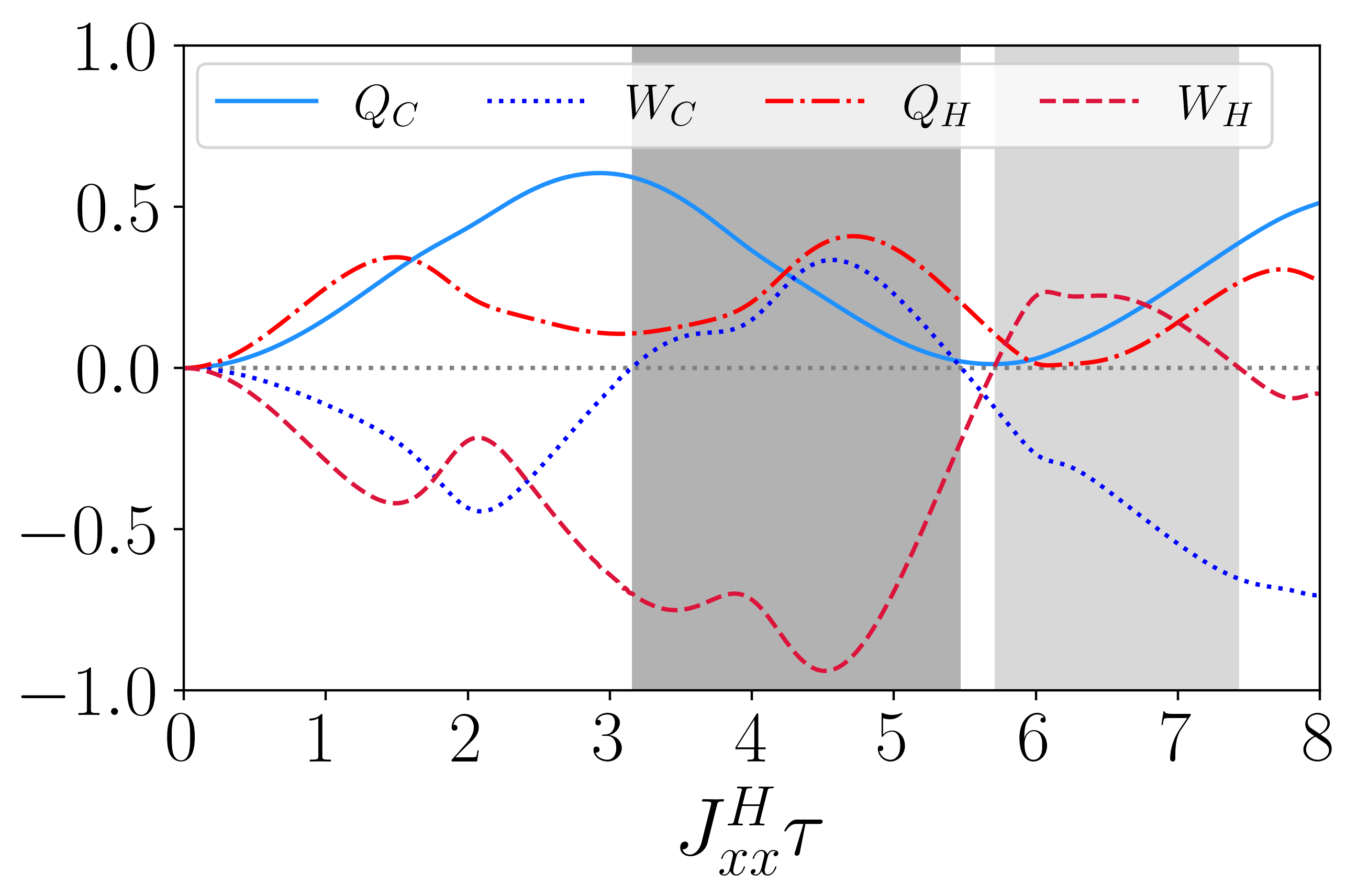}
\caption{Simultaneous-coupling model: Thermodynamical functions at limit cycle as a function of $J_{xx}^{H}\tau$. 
Parameters are $J_{yy}^{H} =\frac{1}{4}J_{xx}^{H},J_{xx}^{C}=\frac{1}{2}J_{xx}^{H},J_{yy}^{C}=\frac{1}{8}J_{xx}^{H}$,  $\tau=0.5$, $\beta_{C}=2$ and $\beta_{H}=1$. The light-grey (dark-grey) region highlights domains where $W_{H}>0$ ($W_{C}>0$).    }
    \label{fig:cutS}
\end{figure}

In the limit of  $J\tau\ll1$, we obtain 
\begin{equation}
\begin{split}
    Q_{C}^{(n+1)} &= \bigg[(J_{xx}^{C}-J_{yy}^{C})^{2}(2p_{C}-1)\\&+4J_{xx}^{C}J_{yy}^{C}\big(p_{C}-p_{S}^{(n)}\big)\bigg]\omega\tau^{2}+O(\tau^{3}).
\end{split}
    \label{eq:Dyson-QCold}
\end{equation}
Similarly, for the hot ancilla we get
\begin{equation}
\begin{split}
    Q_{H}^{(n+1)} &=\bigg[(J_{xx}^{H}-J_{yy}^{H})^{2}(2p_{H}-1)\\&+4J_{xx}^{H}J_{yy}^{H}\big(p_{H}-p_{S}^{(n)}\big)\bigg]\omega\tau^{2}+O(\tau^{3}).
\end{split}
    \label{eq:Dyson-QHot}
\end{equation}
To prove that the system cannot act as a quantum refrigerator, we need to show that in the limit cycle $Q_C^{(\infty)}\geq 0$. 
The first term in Eq. 
(\ref{eq:Dyson-QCold}) 
in the limit cycle
is always nonnegative since $p_C\geq 1/2$. 
Furthermore, in Appendix \ref{AppD} we prove that in this perturbative regime, $p_C\geq p_S^{(\infty)}$; simulations beyond the pertubative regime are included in Appendix \ref{AppC}.
This completes the proof for the impossibility of cooling in the model. While the proof is limited to the perturbative limit, in the short collision time limit the two thermal machine models, alternating and simultaneous, behave identically. Therefore, in that limit, we can rely on the proof of the alternating model for no refrigeration (Sec. \ref{Sec:no_refrigerator}) to assert that the simultaneous model will not support refrigeration in that limit. 
Beyond that, numerical simulations (see for example Fig. \ref{fig:mapS1}) along with an application of optimization approaches indicate that in the simultaneous model heat cannot be extracted from the cold bath, even at strong coupling. 
We thus conjecture that the no-cooling theorem holds for the simultaneous model as well.

\subsubsection{Heat conduction}
Considering a purely heat conducting model, $J_{xx}^A=J_{yy}^A$, and furthermore assuming that all coupling parameters are identical, we get 
\begin{equation}
\begin{split}
    Q^{(n+1)}_{C}(\tau) &= \frac{1}{2} \omega \sin ^2\left(\sqrt{2} J \tau \right)\\&\times \left[\cos \left(2 \sqrt{2} J
   \tau \right) (p_C+p_H-2 p_S^{(n)})\right.\\&\left.+3 p_C-p_H-2
   p_S^{(n)}\right].
\end{split}
\end{equation}
A parallel result is obtained for $Q_H^{(n+1)}$, with
$p_H$ and $p_C$ exchanged.
In the steady state limit, it results that
\begin{equation}
    Q_C^{(\infty)} = -Q_H^{(\infty)} = \omega (p_C-p_H) \sin ^2\left(\sqrt{2} J \tau \right).
\label{eq:Sim_equalJ_Qc_ss}
\end{equation}
Taking the derivative of Eq. (\ref{eq:Sim_equalJ_Qc_ss}) with respect to $\tau$, we get
\begin{equation}
 \dot{Q}_{C}(\tau) = \sqrt{2}\omega J(p_{C}-p_{H}) \sin{\left(2\sqrt{2}J\tau\right)},
\end{equation}
corresponding to
\begin{equation}
    \dot{Q}(y) = 2\sqrt{2}\omega J (p_{C}-p_{H})\sqrt{y(1-y)}\:
\end{equation}
if one defines an effective coupling, $y=\sin^{2}\left(\sqrt{2}J\tau\right)$.
In the weak coupling limit, the heat exchange
is  $Q_C^{(\infty)} = -Q_H^{(\infty)} = 2 \omega (p_C-p_H) (J \tau)^2$, and the heat current is $\dot{Q}_{C}^{(\infty)} =\dot{Q}_{H}^{(\infty)}  =2 \omega (p_C-p_H) \Gamma$. 
We present the heat exchange at each collision in Fig. \ref{fig:Heat1S}, demonstrating an oscillatory behavior. The heat exchange and the heat current are also presented in Fig. \ref{fig:Heat2S} as a function of $J$, with the heat current demonstrating a turnover behavior with $J$.

\subsubsection{Work}
The work component at the hot bath is given by
\begin{equation}
    W_H^{(n+1)} = \text{Tr}\Big[\Big(\hat{U}^{\dagger}(\tau)\hat{H}_{I}^H\hat{U}(\tau)-\hat{H}_{I}\Big)\rho_{S}^{(n)}\otimes\rho_{H}\otimes\rho_{C}\Big]\:.
\end{equation}
An analogue expression can be obtained for the work component at the cold bath, $W_{C}^{(n+1)}$.\\
Simplifying this expression for $J\tau\ll1$ to the second order in $\tau$, we get
\begin{equation}
    W_{H}^{(n+1)}=2(J_{xx}^{H}-J_{yy}^{H})^{2}\omega(1-p_{H}-p_{S}^{(n)})\tau^2.
    \label{eq:Dyson-WHot}
\end{equation}
In the same limit, for the cold bath we get the work 
\begin{equation}
    W_{C}^{(n+1)}=2(J_{xx}^{C}-J_{yy}^{C})^{2}\omega(1-p_{C}-p_{S}^{(n)})\tau^2.
    \label{eq:Dyson-WCold}
\end{equation}
We present numerical results in Fig. \ref{fig:mapS1} with highlighted regimes in Fig. \ref{fig:cutS}. 
Here, we identify two types of performances: Regimes in which the interaction energy is generated at either the hot  $W^{(H,\infty)}\geq 0$,
or cold $W^{(C,\infty)}\geq 0$ contacts. 
However, extensive simulations demonstrated that the total work is always nonpositive, thus, while work can be done locally, overall the system cannot act as a heat engine. 

\begin{figure*}
    \centering
    \includegraphics[width=0.9\linewidth]{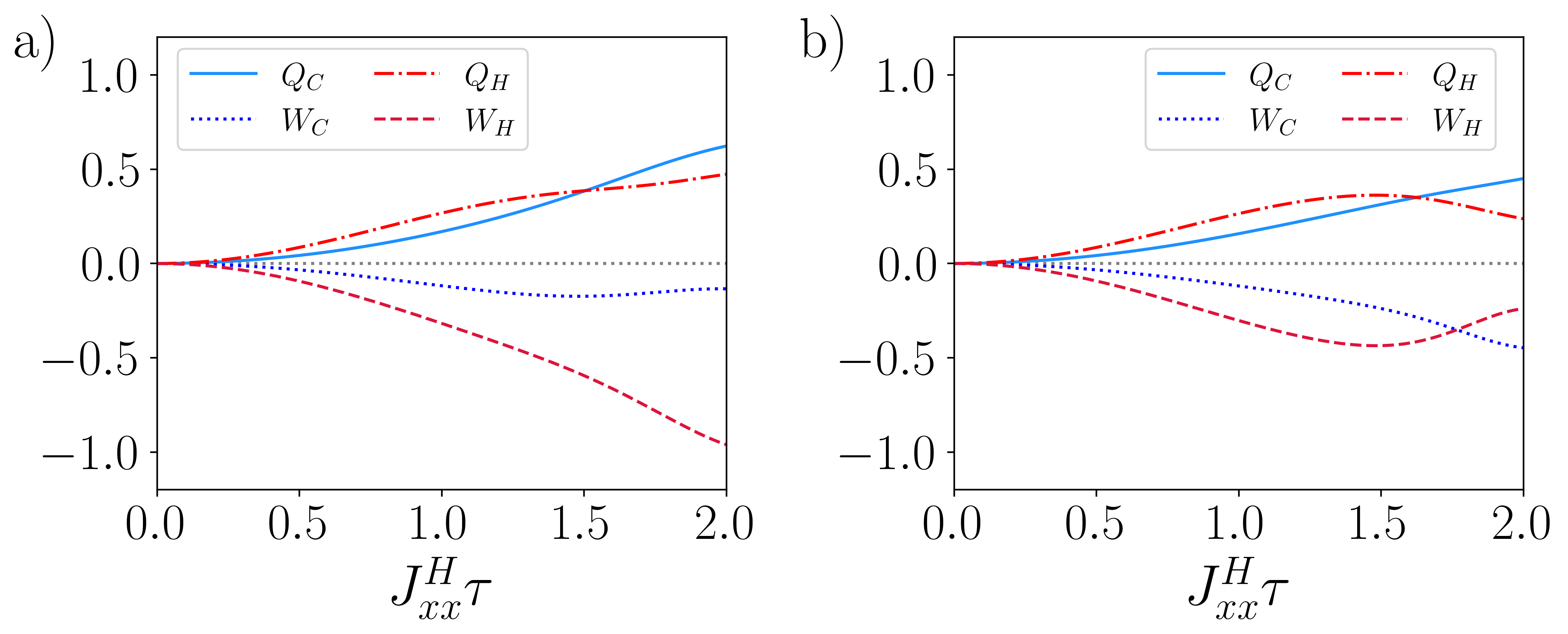}
    \caption{Comparison between (a) alternating and (b) simultaneous thermal machines. We present the thermodynamic functions in the limit cycle as a function of $J_{xx}^{H}\tau$. Other parameters are $J_{yy}^{H} =\frac{1}{4}J_{xx}^{H}$, $J_{xx}^{C}=\frac{1}{2}J_{xx}^{H}$, $J_{yy}^{C}=\frac{1}{8}J_{xx}^{H}$. We set $\tau=0.01$ as collision time, and use $\beta_{C}=2$ and $\beta_{H}=1$ as inverse temperatures of the cold and hot baths' ancillas, respectively.}
    \label{fig:comp}
\end{figure*}

\subsection{Summary of performance: Simultaneous coupling thermal machine}
\label{sec:IIID}

In this section we analyzed a qubit-based thermal machine within the RI framework under \emph{simultaneous} coupling to hot and cold baths. In contrast to the alternating model of Sec. \ref{sec:Alter}, in the simultaneous model each collision involves a three-body interaction between the system and both ancillas. While an exact solution for arbitrary collision times is generally intractable, we obtained analytical results in the so-called Dyson-expansion short-collision-time limit, $J\tau\ll1$, and complemented them with numerical simulations beyond the perturbative regime.

When both $J\tau\ll1$ and $\omega\tau\ll1$, a Trotter argument shows that the two models should behave identically. This can be seen by comparing the behavior of the system, Fig. \ref{fig:popA} to \ref{fig:SimultaneousModel_dynamics}, and their thermodynamical function, Fig. \ref{fig:comp}. We now summarize our observations for the simultaneous model.

{\it I. State of the system.}
One key result is the characterization of the steady-state population reached under simultaneous interactions. Using a Dyson-series expansion to second order in the collision duration, we derived a closed-form expression for the fixed point of the population dynamics, Eq. (\ref{eq:Dyson-sim-ss}). This short-time limit coincides exactly with the limit-cycle population obtained in the alternating model, Eq. (\ref{eq:popAdt}). This establishes a nontrivial equivalence between the two models in the rapid-collision regime.
We showed that, in general, the system does not thermalize to either bath temperatures. Depending on the asymmetry between the coupling parameters, the limit-cycle population can lie outside the interval defined by the hot and cold bath populations, corresponding to an effective temperature higher than both baths. This ``overheating'' effect is directly linked to work performed on the system through the switching of the interaction Hamiltonian and requires anisotropic couplings, $J_{xx} \neq J_{yy}$.

{\it II. Refrigeration.}
The thermodynamic analysis revealed that the simultaneous-coupling model obeys the same fundamental constraint as the alternating model: refrigeration is impossible. In the perturbative limit, we proved analytically that the heat exchanged with the cold bath in the limit cycle is always nonnegative, $Q_C^{(\infty)} \ge 0$, ruling out heat extraction from the cold reservoir. Numerical simulations support the conjecture that this result persists beyond the short-time regime and at strong coupling.

{\it III. Heat conduction.}
In the absence of work contributions, the model reduces to a purely heat-conducting junction. As in the alternating case, the heat exchanged per collision displays oscillatory behavior as functions of the coupling strength and the collision time. Once again the heat current exhibits a turnover behavior, increasing at weak coupling and becoming suppressed at strong coupling, consistent with the emergence of effective decoupling in the ultrastrong-coupling regime.

{\it IV. Work contributions. }
We examined work generation in the simultaneous model. While positive work can be generated locally at one of the contacts under suitable parameter choices, the total work performed over a collision remained nonpositive in the regimes explored. As a result, the simultaneous-coupling setup cannot operate as a heat engine. 

Our results demonstrate that allowing simultaneous system-bath interactions does not circumvent the fundamental limitations identified in Sec. \ref{sec:Alter}, and further underscores the necessity of extending beyond minimal qubit-based, Markovian repeated-interaction models to realize functional quantum thermal machines.

\section{Summary}
\label{sec:Summ}
We analyzed the performance of a qubit functioning as a quantum thermal device within the repeated interaction formalism, which allowed access to regimes beyond standard weak-coupling and stroboscopic descriptions. Two models were studied: an alternating-coupling setup, where the qubit interacted sequentially with hot and cold ancillas, and a simultaneous-coupling setup, where it interacted with both baths at once. For the alternating model, we derived an exact analytical solution for the population dynamics and demonstrated convergence to a limit cycle. A central result was the rigorous no-go theorem for quantum refrigeration: regardless of coupling strength, interaction time, or parameter asymmetry, the qubit cannot extract heat from the cold bath. This establishes a fundamental constraint on qubit-based thermal machines operating under simple RI dynamics.
This impossibility of refrigeration was further corroborated in the simultaneous-coupling model, where a perturbative analysis showed that heat extraction from the cold reservoir was forbidden, reinforcing the generality of the constraint within RI-based qubit machines.

Beyond refrigeration, our study provided a detailed thermodynamic analysis of work and heat in the model. In the alternating-coupling model, while work can be generated locally at one contact (the hot bath interface), the total work over a full cycle is always nonpositive, implying that the device cannot operate as a genuine heat engine. In regimes without work exchange, the model exhibited purely heat conductive behavior, with heat flowing from hot to cold in a manner that shows strong-coupling signatures, including oscillatory suppression and revival of heat currents as a function of the interaction strength. These features persisted across both alternating and simultaneous interaction schemes, the latter analyzed perturbatively, reinforcing the robustness of the identified thermodynamic limitations.

Our results show that a minimal qubit working fluid coupled to memoryless, uncorrelated ancillas is insufficient for realizing functional quantum refrigerators or engines within the RI framework. This points toward several extensions: incorporating non-Markovian bath structures, allowing correlations between ancillas, introducing coherence or squeezing in the reservoirs, enlarging the working fluid to multilevel systems, or separating the work and the heat strokes. Such generalizations, some discussed previously, can relax the no-go constraints identified here and enable quantum thermal machines. More broadly, our study positions the repeated interaction formalism as a powerful platform for isolating which physical ingredients are necessary to realize useful quantum thermal machines.


\begin{acknowledgments}
We acknowledge fruitful discussions with David Gelbwaser-Klimovsky on the turnover behavior in heat exchange at strong coupling and on different types of collisional models, inspiring our analysis of this phenomenon.
DS acknowledges support from an NSERC Discovery Grant.
The work of MG was supported by the NSERC Canada Graduate Scholarship-Doctoral. 
GBG and AP were supported by the research project: ``Quantum Software Consortium: Exploring Distributed Quantum Solutions for Canada" (QSC). QSC is financed under the National Sciences and Engineering Research Council of Canada (NSERC) Alliance Consortia Quantum Grants \#ALLRP587590-23. 
GBG summer project was further supported through an NSERC USRA award.
\end{acknowledgments}

\appendix
\begin{widetext}
    
\section{Proof that $p_S^{(\infty,A)}\leq p_C$ in the alternating-coupling thermal machine}
\label{AppA}

We want to show that in the limit cycle, after a collision with the cold bath, the ground state population of the system is upper bounded by the population of a cold ancilla,
\bea
p_S^{(\infty, C)} \leq p_C.
\eea
In other words, the system cannot be made colder than the cold bath.
First, we prove that the denominator of Eq. (\ref{eq:pop}) is positive (see also Ref. \cite{Prositto2025Dynamics}),
\bea
1-(1-\kappa_{\theta}^C-\kappa_{\phi}^C) (1-\kappa_{\theta}^H-\kappa_{\phi}^H)  > 0
\eea
Note that we do not consider the special points at which the dynamics is frozen or all couplings are zero. 
Opening the products, we get
\bea
-\kappa_{\theta}^H
-\kappa_{\phi}^H
-\kappa_{\theta}^C
+\kappa_{\theta}^C\kappa_{\theta}^H
+\kappa_{\theta}^C\kappa_{\phi}^H
-\kappa_{\phi}^C
+\kappa_{\phi}^C\kappa_{\theta}^H
+\kappa_{\phi}^C\kappa_{\phi}^H <0
\eea
All these terms are bounded, 
$0<\kappa_{\theta}^A<1$
and $0\leq\kappa_{\phi}^A<1$.
We now note that
\bea
\kappa_{\theta}^H>\kappa_{\theta}^H \kappa_{\phi}^C, \,\,\,\,
\kappa_{\phi}^H>\kappa_{\phi}^H \kappa_{\theta}^C,\,\,\,\,
\kappa_{\theta}^C >\kappa_{\theta}^C \kappa_{\theta}^H, \,\,\,\,
\kappa_{\phi}^C >\kappa_{\phi}^C  \kappa_{\phi}^H;
\eea
which completes the proof.

We now return to Eq. (\ref{eq:pop}) and reorganize it
\bea
 \kappa_{\theta}^C p_C+\kappa_{\phi}^C (1-p_C)+(1-\kappa_{\theta}^C-\kappa_{\phi}^C) [\kappa_{\theta}^H p_H+\kappa_{\phi}^H (1-p_H)]
 \leq p_C-p_C(1-\kappa_{\theta}^C-\kappa_{\phi}^C) (1-\kappa_{\theta}^H-\kappa_{\phi}^H)\:.
\eea
We rearrange the inequality (right side) as follows
\bea
&& [\kappa_{\theta}^C p_C+\kappa_{\phi}^C (1-p_C)]+(1-\kappa_{\theta}^C-\kappa_{\phi}^C) \kappa_{\theta}^H p_H +(1-\kappa_{\theta}^C-\kappa_{\phi}^C) 
\kappa_{\phi}^H (1-p_H)
 \nonumber\\
&&\leq [p_C -p_C(1-\kappa_{\theta}^C-\kappa_{\phi}^C) ]
 +(1-\kappa_{\theta}^C-\kappa_{\phi}^C) \kappa_{\theta}^Hp_C
  +(1-\kappa_{\theta}^C-\kappa_{\phi}^C) \kappa_{\phi}^Hp_C\:.
  \label{eq:A1}
\eea
We now compare the first term in the left side of inequality (\ref{eq:A1}) to the first term in the right side of that relation. We can easily confirm that
\bea
[\kappa_{\theta}^C p_C+\kappa_{\phi}^C (1-p_C)] 
\leq 
[p_C -p_C(1-\kappa_{\theta}^C-\kappa_{\phi}^C) ],
\eea
since $p_C\geq 1/2$.
Next, we compare the second term in the left side of inequality (\ref{eq:A1}) to the second term in the right side of that expression, showing that
\bea
(1-\kappa_{\theta}^C-\kappa_{\phi}^C) \kappa_{\theta}^H p_H  \leq 
(1-\kappa_{\theta}^C-\kappa_{\phi}^C) \kappa_{\theta}^Hp_C,
\eea
since $p_C\geq p_H$.
Lastly, we compare the third term in the left side of inequality (\ref{eq:A1}) to the third term in the right side of the  inequality, noting that
\bea
(1-\kappa_{\theta}^C-\kappa_{\phi}^C) 
\kappa_{\phi}^H (1-p_H)
\leq (1-\kappa_{\theta}^C-\kappa_{\phi}^C) \kappa_{\phi}^Hp_C,
\eea
since $1-p_H\leq p_C$. This concludes our proof of the validity of (\ref{eq:A1}).
Together, we have proved the original statement, $p_S^{(\infty, C)} \leq p_C$: in the alternating model the cold ancilla's population acts as an upper bound on the system's ground-state population in the limit cycle.
Following similar steps, one can prove that
\bea
p_S^{(\infty, H)}\leq p_C,
\eea 
namely, the system's ground state population after a collision with the hot bath is always upper bounded by the temperature of the cold bath.
This is once again intuitive, as we do not expect the hot bath to cool the system below the temperature of the cold bath. On the other hand, while
one may naively expect the system to be hotter after a collision with a hot ancilla, than after a collision with a cold ancilla, that is
$p_S^{(\infty,H)}\leq p_S^{(\infty,C)}$, this relation is incorrect, as we demonstrate with simulations in Fig. \ref{fig:figpss}(c).

\section{Proof that $1-p_{C}\leq p_{S}^{(\infty,H)}$ in the alternating-coupling model}
\label{AppB}

Our goal is to show that in the alternating model $p_{S}^{(\infty,H)}\geq1-p_{C}$. From Eq. (\ref{eq:pop}), we have
\begin{equation}
     \frac{\kappa_{\theta}^H p_H+\kappa_{\phi}^H (1-p_H)+(1-\kappa_{\theta}^H-\kappa_{\phi}^H) [\kappa_{\theta}^C p_C+\kappa_{\phi}^C (1-p_C)]}{1-(1-\kappa_{\theta}^C-\kappa_{\phi}^C) (1-\kappa_{\theta}^H-\kappa_{\phi}^H)}-(1-p_{C})\geq0 \:.
\label{eq:LowerBound_alternating}     
\end{equation}
First, we note that $\kappa_{\alpha}^{A}$ with $A=C,H$ and $\alpha=\theta,\phi$ is always non-negative (see Eq. (\ref{eq:kappa})). Second, we observe that $1/2\leq p_{A}\leq1$ implies $0\leq 1-p_{A}\leq1/2$. We can thus state that $1-p_{A}\leq p_{A}$, for $A=C,H$.
Therefore, for the first term in the right-hand-side of Eq. (\ref{eq:LowerBound_alternating}), we write 
\begin{equation}
\frac{\kappa_{\theta}^H p_H+\kappa_{\phi}^H (1-p_H)+(1-\kappa_{\theta}^H-\kappa_{\phi}^H) [\kappa_{\theta}^C p_C+\kappa_{\phi}^C (1-p_C)]}{1-(1-\kappa_{\theta}^C-\kappa_{\phi}^C) (1-\kappa_{\theta}^H-\kappa_{\phi}^H)}\geq \frac{(\kappa_{\theta}^{H}+\kappa_{\phi}^{H})(1-p_{H})+(\kappa_{\theta}^{C}+\kappa_{\phi}^{C})(1-p_{C})(1-\kappa_{\theta}^{H}-\kappa_{\phi}^{H})}{1-(1-\kappa_{\theta}^C-\kappa_{\phi}^C) (1-\kappa_{\theta}^H-\kappa_{\phi}^H)}\:.
\end{equation}
To prove Eq. (\ref{eq:LowerBound_alternating}), it is thus sufficient to show that
\begin{equation}
    \frac{(\kappa_{\theta}^{H}+\kappa_{\phi}^{H})(1-p_{H})+(\kappa_{\theta}^{C}+\kappa_{\phi}^{C})(1-p_{C})(1-\kappa_{\theta}^{H}-\kappa_{\phi}^{H})}{1-(1-\kappa_{\theta}^C-\kappa_{\phi}^C) (1-\kappa_{\theta}^H-\kappa_{\phi}^H)}-(1-p_{C})\geq0\:,
\end{equation}
namely
\begin{equation}
\frac{(\kappa_{\theta}^{H}+\kappa_{\phi}^{H})(1-p_{H})+(\kappa_{\theta}^{C}+\kappa_{\phi}^{C})(1-p_{C})(1-\kappa_{\theta}^{H}-\kappa_{\phi}^{H})-[1-(1-\kappa_{\theta}^C-\kappa_{\phi}^C) (1-\kappa_{\theta}^H-\kappa_{\phi}^H)](1-p_{C})}{1-(1-\kappa_{\theta}^C-\kappa_{\phi}^C) (1-\kappa_{\theta}^H-\kappa_{\phi}^H)}\geq0\:.
\label{eq:Lowerbound_alternating2}
\end{equation}
Simplifying Eq. (\ref{eq:Lowerbound_alternating2}) and noting that the denominator is non-negative (see Appendix \ref{AppA}),
we remain with  
\begin{equation}
(\kappa_{\theta}^{H}+\kappa_{\phi}^{H})(1-p_{H})+(\kappa_{\theta}^{C}+\kappa_{\phi}^{C})(1-p_{C})(1-\kappa_{\theta}^{H}-\kappa_{\phi}^{H})-[1-(1-\kappa_{\theta}^H-\kappa_{\phi}^H)+ (1-\kappa_{\theta}^H-\kappa_{\phi}^H)(\kappa_{\theta}^C+\kappa_{\phi}^C)](1-p_{C})\geq0\:,
\end{equation}
namely,
\begin{equation}
(\kappa_{\theta}^H+\kappa_{\phi}^H)(p_{C}-p_{H})\geq0\:.
\end{equation}
This is always true since $p_{C}\geq p_{H}$, proving our original statement, Eq. (\ref{eq:LowerBound_alternating}).


\begin{figure}[htbp]
    \centering
\includegraphics[width=1.0\linewidth]{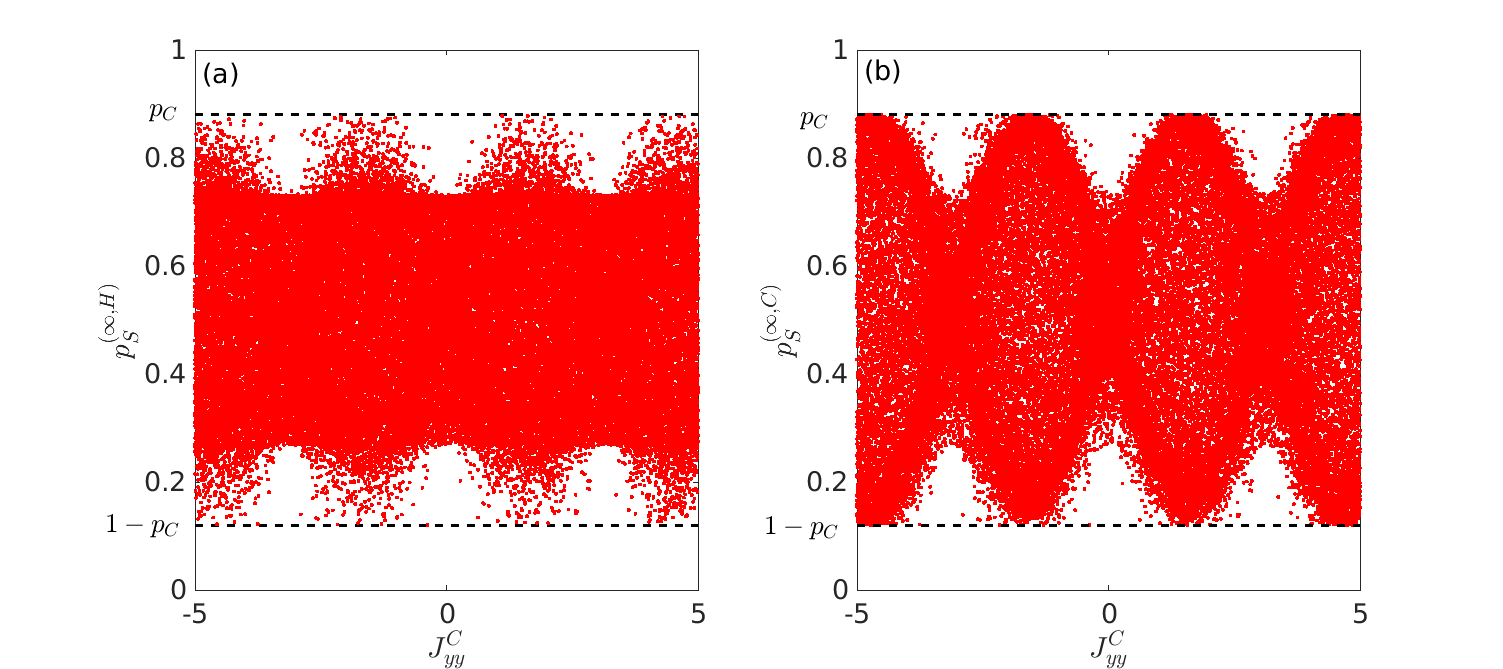}
    \caption{Alternating-coupling model. (a) Numerical example for the existence of lower and upper bounds on the system's population after a collision with a hot ancilla, $1-p_C \leq p_S^{(\infty, H)}\leq p_C$.
    (b) A parallel bound holds for 
    the population after a collision with a cold ancilla, $1-p_C\leq p_S^{(\infty, C)}\leq p_C$.
    Parameters are $\tau=0.5$, $\beta_H=1$, $\beta_C=2$. Results here were generated with 60,000 samples of the four coupling parameters, each independently varied and selected from a uniform distributions in the range -5 to +5. The x-axis as such could have been any of the coupling coefficients. Results were generated using the analytical expressions for the population at limit cycle, Eq. (\ref{eq:pop}).}
    \label{fig:bounds}
\end{figure}

\begin{figure}[htbp]
    \centering
\includegraphics[width=0.5\linewidth]{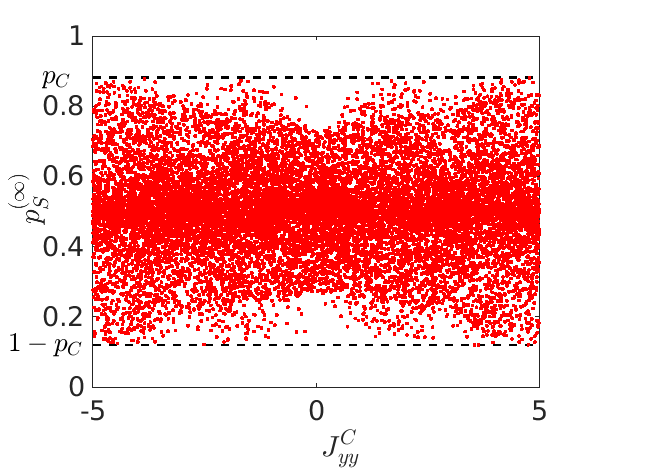}
    \caption{Simultaneous-coupling model: (a) Numerical support for the existence of lower and upper bounds on the system's population after a collision with two ancillas, hot and cold, $1-p_C \leq p_S^{(\infty)}\leq p_C$.
    Parameters are $\tau=0.5$, $\beta_H=1$, $\beta_C=2$. Results here were generated from 15,000 samples of the four coupling parameters, each independently varied and selected from a uniform distributions in the range -5 to +5. The x-axis as such could have been any of the coupling coefficients. Results were generated from brute-force numerical time evolution.}
    \label{fig:boundsS}
\end{figure}
\section{Numerical support for population bounds
in both alternating-coupling and simultaneous-coupling thermal machines}
\label{AppC}

We provide here numerical demonstrations of the bounds on the limit cycle population. While for the alternating-coupling model we derived rigorous upper and lower bounds on the population in Appendices \ref{AppA} and \ref{AppB}, respectively, in the simultaneous model we did not obtain such proofs for the general case, but only in the stroboscopic limit, which trivially reduces to the alternating case.
For the simultaneous model, we thus miss such a rigorous proof for the case with strong interactions and long collisions. 

\subsection{Alternating-coupling model}
We are interested in demonstrating that the population of the system after a collision with a hot bath is both lower and upper bounded,
\bea 
1-p_C \leq p_S^{(\infty, H)}\leq p_C.
\label{eq:bounds}
\eea
The population is lower bounded by the cold ancilla's population, meaning that we cannot cool the system below the cold bath's temperature, as derived in Appendix \ref{AppA}. As for the lower bound, it points to the fact that one cannot arbitrarily overheat the system. The hottest it can get is bounded by population inversion with respect to the cold bath. We proved this bound analytically in Appendix \ref{AppB}.

In Fig. \ref{fig:bounds}(a) we study the limit cycle population of the system after a collision with a hot ancilla. We repeat these calculations 60,000 times, each time independently sampling from a uniform distribution each of the four coupling parameters from the range of -5 to +5. As we see in Fig. \ref{fig:bounds}(a), the limit cycle population is clearly bounded and satisfying Eq. (\ref{eq:bounds}).
As the number of samples increases, the region between the bounds becomes fully covered.

In Fig. \ref{fig:bounds}(b) we probe bounds
on the population of the system after a collision with a cold ancilla,
\bea 
1-p_C \leq p_S^{(\infty, C)}\leq p_C.
\label{eq:boundsC}
\eea
Once again, the upper bound reflects that the system cannot be made colder than the cold ancilla. The lower bound is more subtle, and it reflects that the system cannot be made ``too" hot, and it is bounded by what population inversion with respect to the cold ancilla allows.  

\subsection{Simultaneous-coupling model}

In Fig. \ref{fig:boundsS} we demonstrate bounds on limit cycle population for a qubit coupled simultaneously to two thermal baths,
\bea 
1-p_C \leq p_S^{(\infty)}\leq p_C.
\label{eq:boundsS}
\eea
This example adopts a collision time of $\tau=0.5$, which extends the analytical limit probed in Sec. \ref{sec:Simult}.
Since we do not have a general analytical expression for the limit cycle, valid at strong coupling and long collision time, simulations here were done by performing a time evolution from a certain initial condition for each combination of interaction parameters. 

\section{Proof that $p_S^{(\infty)}\leq p_C$ in the  simultaneous-coupling Model under perturbative $J\tau\ll1$ couplings}
\label{AppD}

Considering the simultaneous model and its perturbative solution for the system ground state population at limit cycle, Eq. (\ref{eq:Dyson-sim-ss}), we prove here that 
\bea
p_S^{(\infty)}\leq p_C.
\label{eq:C1}
\eea
This inequality (\ref{eq:C1}) corresponds to
\bea
\frac{\left(J_{xx}^{C}-J_{yy}^{C}\right)^{2}+\left(J_{xx}^{H}-J_{yy}^{H}\right)^{2} +4J_{xx}^{C}J_{yy}^{C}p_{C}+4J_{xx}^{H}J_{yy}^{H}p_{H}}{2\left[\left(J_{xx}^{H}\right)^{2}+\left(J_{yy}^{H}\right)^{2}+\left(J_{xx}^{C}\right)^{2}+\left(J_{yy}^{C}\right)^{2}\right]} \leq p_C \:.
\label{eq:UpperBound_simultaneous} 
\eea
Since $p_{H}\leq p_{C}$, we can write
\begin{equation}
\frac{\left(J_{xx}^{C}-J_{yy}^{C}\right)^{2}+\left(J_{xx}^{H}-J_{yy}^{H}\right)^{2} +4J_{xx}^{C}J_{yy}^{C}p_{C}+4J_{xx}^{H}J_{yy}^{H}p_{H}}{2\left[\left(J_{xx}^{H}\right)^{2}+\left(J_{yy}^{H}\right)^{2}+\left(J_{xx}^{C}\right)^{2}+\left(J_{yy}^{C}\right)^{2}\right]} 
\leq 
\frac{\left(J_{xx}^{C}-J_{yy}^{C}\right)^{2}+\left(J_{xx}^{H}-J_{yy}^{H}\right)^{2} +4J_{xx}^{C}J_{yy}^{C}p_{C}+4J_{xx}^{H}J_{yy}^{H}p_{C}}{2\left[\left(J_{xx}^{H}\right)^{2}+\left(J_{yy}^{H}\right)^{2}+\left(J_{xx}^{C}\right)^{2}+\left(J_{yy}^{C}\right)^{2}\right]}\:.
\end{equation}
To prove the inequality in Eq. (\ref{eq:UpperBound_simultaneous}) it is sufficient to show that 
\begin{equation}
\frac{\left(J_{xx}^{C}-J_{yy}^{C}\right)^{2}+\left(J_{xx}^{H}-J_{yy}^{H}\right)^{2} +4J_{xx}^{C}J_{yy}^{C}p_{C}+4J_{xx}^{H}J_{yy}^{H}p_{C}}{2\left[\left(J_{xx}^{H}\right)^{2}+\left(J_{yy}^{H}\right)^{2}+\left(J_{xx}^{C}\right)^{2}+\left(J_{yy}^{C}\right)^{2}\right]}-p_{C}\leq0\:.
\label{eq:Upperbound_simultaneous2}
\end{equation}
Simplifying Eq. (\ref{eq:Upperbound_simultaneous2}) and noting that the denominator is positive-defined, we get
\begin{equation}
\Big[(J_{xx}^{H}-J_{yy}^{H})^{2}+(J_{xx}^{C}-J_{yy}^{C})^{2}\Big](1-2p_{C})\leq0,
\end{equation}
which is always true since $p_{C}\geq1/2$. This proves our original statement, Eq. (\ref{eq:C1}).
\end{widetext}

\bibliography{references}

@ARTICLE{Nayem2025,
  title     = "Nanotechnology and thermodynamics: a comprehensive review",
  author    = "Rahman, Mustafizur and Rimon, Md Israfil Hossain and Hasan,
               Zahid and Oliullah, Md Shah and Rabbi, S M Fazle and Mobarak, Md
               Hosne and Hossain, Nayem",
  journal   = "Nanomater. Energy",
  publisher = "Emerald",
  volume    =  14,
  number    =  2,
  pages     = "109--134",
  month     =  jun,
  year      =  2025,
  copyright = "https://creativecommons.org/licences/by/4.0/",
  url= "https://doi.org/10.1680/jnaen.24.00093"
}

@article{Auffeves2022,
  title = {Quantum Technologies Need a Quantum Energy Initiative},
  author = {Auff\`eves, Alexia},
  journal = {PRX Quantum},
  volume = {3},
  issue = {2},
  pages = {020101},
  numpages = {12},
  year = {2022},
  month = {Jun},
  publisher = {American Physical Society},
  doi = {10.1103/PRXQuantum.3.020101},
  url = {https://link.aps.org/doi/10.1103/PRXQuantum.3.020101}
}

@ARTICLE{Dutta2021,
  title     = "Quantum thermal machines and batteries",
  author    = "Bhattacharjee, Sourav and Dutta, Amit",
  journal   = "Eur. Phys. J. B",
  publisher = "Springer Science and Business Media LLC",
  volume    =  94,
  pages = 239,
  number    =  12,
  month     =  dec,
  year      =  2021,
  copyright = "https://www.springernature.com/gp/researchers/text-and-data-mining",
  url= "https://doi.org/10.1140/epjb/s10051-021-00235-3"
}

@ARTICLE{Arrachea2025,
  title         = "Lindbladian approach for many-qubit thermal machines:
                   enhancing the performance with geometric heat pumping by
                   entanglement",
  author        = "Caselli, Ger{\'o}nimo J and Manuel, Luis O and Arrachea,
                   Liliana",
  journal = " ",                 
  month         =  nov,
  year          =  2025,
  copyright     = "http://creativecommons.org/licenses/by/4.0/",
  archivePrefix = "arXiv",
  primaryClass  = "quant-ph",
  eprint        = "2511.16591",
  url = "https://doi.org/10.48550/arXiv.2511.16591"
}

@article{Maslennikov2019,
  author  = {Maslennikov, G. and Ding, S. and Habl{\"u}tzel, R. and Gan, J. and Roulet, A.
             and Nimmrichter, S. and Dai, J. and Scarani, V. and Matsukevich, D.},
  title   = {Quantum absorption refrigerator with trapped ions},
  journal = {Nature Communications},
  volume  = {10},
  pages   = {202},
  year    = {2019},
  url = "https://doi.org/10.1038/s41467-018-08090-0"
}

@article{Rossnagel2016,
  author  = {Rossnagel, J. and Dawkins, S. T. and Tolazzi, K. N. and Abah, O.
             and Lutz, E. and Schmidt-Kaler, F. and Singer, K.},
  title   = {A single-atom heat engine},
  journal = {Science},
  volume  = {352},
  pages   = {325--329},
  year    = {2016},
  url = "https://doi.org/10.1126/science.aad6320"
}

@article{Josefsson2018,
  author  = {Josefsson, M. and Svilans, A. and Burke, A. M. and Hoffman, E. A.
             and Fahlvik, S. and Thelander, C. and Leijnse, M. and Linke, H.},
  title   = {A quantum-dot heat engine operating close to the thermodynamic efficiency limits},
  journal = {Nature Nanotechnology},
  volume  = {13},
  pages   = {920--924},
  year    = {2018},
  url = "https://doi.org/10.1038/s41565-018-0200-5"
}

@article{Aamir2025,
  author  = {Aamir, M. A. and Suria, P. J. and Guzman, J. A. M. and Castillo-Moreno, C.
             and Epstein, J. M. and Halpern, N. Y. and Gasparinetti, S.},
  title = "Thermally driven quantum refrigerator autonomously resets a superconducting qubit",
  journal = {Nature Physics},
  volume  = {21},
  pages   = {318--323},
  year    = {2025},
  url = "https://doi.org/10.1038/s41567-024-02708-5"
}

@article{Klaers2017,
  author  = {Klaers, J. and Faelt, S. and Imamo{\u g}lu, A. and Togan, E.},
  title   = {Squeezed thermal reservoirs as a resource for a nanomechanical engine beyond the {C}arnot limit},
  journal = {Physical Review X},
  volume  = {7},
  pages   = {031044},
  year    = {2017},
  url = "https://doi.org/10.1103/PhysRevX.7.031044"
}

@article{Ahmadi2025,
  author  = {Ahmadi, Z. and Mojaveri, B.},
  title   = {Efficiency enhancement up to unity in a generalized quantum Otto engine:
             Comparative analysis with conventional quantum Otto engine utilizing a two-qubit Heisenberg XXZ chain},
  journal = {Physical Review B},
  volume  = {112},
  pages   = {014116},
  year    = {2025},
  url = "https://doi.org/10.1103/kg8j-yv1n"
}

@article{Prositto2025Dynamics,
  author  = {Prositto, A. and Forbes, M. and Segal, D.},
  title   = {Equilibrium and nonequilibrium steady states with the repeated interaction protocol:
             relaxation dynamics and energetic cost},
  journal = {Quantum Science and Technology},
  volume  = {10},
  pages   = {025061},
  year    = {2025},
  url = "https://doi.org/10.1088/2058-9565/adc7d4"
}

@article{RamonEscandel2025,
  title = {Thermal state preparation by repeated interactions at and beyond the Lindblad limit},
  author = {Ramon-Escandell, Carlos and Prositto, Alessandro and Segal, Dvira},
  journal = {Phys. Rev. Res.},
  volume = {7},
  issue = {4},
  pages = {043012},
  numpages = {35},
  year = {2025},
  month = {Oct},
  publisher = {American Physical Society},
  doi = {10.1103/5y2q-jzgx},
  url = {https://link.aps.org/doi/10.1103/5y2q-jzgx}
}

@article{RomanAncheyta2021,
  author  = {Rom{\'a}n-Ancheyta, R. and Kol{\'a}{\v r}, M. and Guarnieri, G. and Filip, R.},
  title   = {Enhanced steady-state coherence via repeated system-bath interactions},
  journal = {Physical Review A},
  volume  = {104},
  pages   = {062209},
  year    = {2021},
  url = "https://doi.org/10.1103/PhysRevA.104.062209"
}

@article{Segal05,
  author  = {Segal, D. and Nitzan, A.},
  title   = {Spin-boson thermal rectifier},
  journal = {Physical Review Letters},
  volume  = {94},
  pages   = {034301},
  year    = {2005},
  url = "https://doi.org/10.1103/PhysRevLett.94.034301"
}

@article{Nicolin11,
  author  = {Nicolin, L. and Segal, D.},
  title   = {Non-equilibrium spin-boson model: Counting statistics and the heat exchange fluctuation theorem},
  journal = {The Journal of Chemical Physics},
  volume  = {135},
  pages   = {164106},
  year    = {2011},
  url = "https://doi.org/10.1063/1.3655674"
}

@article{Anto21,
  author  = {Anto-Sztrikacs, N. and Segal, D.},
  title   = {Strong coupling effects in quantum thermal transport with the reaction coordinate method},
  journal = {New Journal of Physics},
  volume  = {23},
  pages   = {063036},
  year    = {2021},
  url = "https://doi.org/10.1088/1367-2630/ac02df"
}

@article{Kilgour19,
  author  = {Kilgour, M. and Agarwalla, B. K. and Segal, D.},
  title   = {Path-integral methodology and simulations of quantum thermal transport:
             Full counting statistics approach},
  journal = {The Journal of Chemical Physics},
  volume  = {150},
  pages   = {084107},
  year    = {2019},
  url = "https://doi.org/10.1063/1.5084949"
}

@article{Agarwalla17,
  author  = {Agarwalla, B. K. and Segal, D.},
  title   = {Energy current and its statistics in the nonequilibrium spin-boson model:
             Majorana fermion representation},
  journal = {New Journal of Physics},
  volume  = {19},
  pages   = {043030},
  year    = {2017},
  url = "https://doi.org/10.1088/1367-2630/aa6657"
}

@article{Segal14,
  title = {Heat transfer in the spin-boson model: A comparative study in the incoherent tunneling regime},
  author = {Segal, Dvira},
  journal = {Phys. Rev. E},
  volume = {90},
  issue = {1},
  pages = {012148},
  numpages = {6},
  year = {2014},
  month = {Jul},
  publisher = {American Physical Society},
  doi = {10.1103/PhysRevE.90.012148},
  url = {https://link.aps.org/doi/10.1103/PhysRevE.90.012148}
}

@article{Nazim14,
  author    = {N. Boudjada and D. Segal},
  title     = {From dissipative dynamics to studies of heat transfer at the nanoscale: Analysis of the spin-boson model},
  journal   = {The Journal of Physical Chemistry A},
  year      = {2014},
  volume    = {118},
  number    = {47},
  pages     = {11323--11336},
  url = "https://doi.org/10.1021/jp5091685"
}

@article{Thoss10,
    author = {Velizhanin, Kirill A. and Thoss, Michael and Wang, Haobin},
    title = {Meir–Wingreen formula for heat transport in a spin-boson nanojunction model},
    journal = {The Journal of Chemical Physics},
    volume = {133},
    number = {8},
    pages = {084503},
    year = {2010},
    month = {08},

    doi = {10.1063/1.3483127},
    url = {https://doi.org/10.1063/1.3483127}
}

@article{Thoss08,
  author  = {Velizhanin, K. A. and Wang, H. and Thoss, M.},
  title   = {Heat transport through model molecular junctions: A multilayer multiconfiguration time-dependent Hartree approach},
  journal = {Chemical Physics Letters},
  volume  = {460},
  pages   = {325--?},
  year    = {2008},
  doi={doi = {https://doi.org/10.1016/j.cplett.2008.05.065},}
}

@article{Cao15,
  author  = {Wang, C. and Ren, J. and Cao, J.},
  title   = {Nonequilibrium Energy Transfer at Nanoscale: A Unified Theory from Weak to Strong Coupling},
  journal = {Scientific Reports},
  volume  = {5},
  pages   = {11787},
  year    = {2015},
  doi={https://doi.org/10.1038/srep11787}
}

@article{Cao17,
  title = {Unifying quantum heat transfer in a nonequilibrium spin-boson model with full counting statistics},
  author = {Wang, Chen and Ren, Jie and Cao, Jianshu},
  journal = {Phys. Rev. A},
  volume = {95},
  issue = {2},
  pages = {023610},
  numpages = {10},
  year = {2017},
  month = {Feb},
  publisher = {American Physical Society},
  doi = {10.1103/PhysRevA.95.023610},
  url = {https://link.aps.org/doi/10.1103/PhysRevA.95.023610}
}

@article{Barra2015,
  title={The thermodynamic cost of driving quantum systems by their boundaries},
  author={Barra, F.},
  journal={Scientific reports},
  volume={5},
  number={1},
  pages={14873},
  year={2015},
  doi={https://doi.org/10.1038/srep14873
},
  publisher={Nature Publishing Group UK London}
}

@article{Xia2019,
  title={The effects of system–environment correlations on heat transport and quantum entanglement via collision models},
  author={Z.-X. Man and Q. Zhang and Y.-J. Xia},
  journal={Quantum Information Processing},
  volume={18},
  number={5},
  pages = {157},
  year={2019},
  publisher={Springer Nature},
  url = "https://doi.org/10.1007/s11128-019-2275-9"
}

@article{DeChiara2024,
  title = {Quantum coherence enables hybrid multitask and multisource regimes in autonomous thermal machines},
  author = {Hammam, Kenza and Manzano, Gonzalo and De Chiara, Gabriele},
  journal = {Phys. Rev. Res.},
  volume = {6},
  issue = {1},
  pages = {013310},
  numpages = {17},
  year = {2024},
  month = {Mar},
  publisher = {American Physical Society},
  doi = {10.1103/PhysRevResearch.6.013310},
  url = {https://link.aps.org/doi/10.1103/PhysRevResearch.6.013310}
}

@ARTICLE{Brunner2024,
  title     = "Thermodynamic computing via autonomous quantum thermal machines",
  author    = "Lipka-Bartosik, Patryk and Perarnau-Llobet, Mart{\'\i} and
               Brunner, Nicolas",
  journal   = "Sci. Adv.",
  publisher = "American Association for the Advancement of Science (AAAS)",
  volume    =  10,
  number    =  36,
  pages     = "eadm8792",
  month     =  sep,
  year      =  2024,
  url = "https://doi.org/10.1126/sciadv.adm8792"
}

@BOOK{Petruccione2002,
  title     = "The theory of open quantum systems",
  author    = "Breuer, Heinz-Peter and Petruccione, Francesco",
  publisher = "Oxford University Press",
  month     =  jan,
  year      =  2007
}

@ARTICLE{Martin2010,
  title     = "Markovian master equations: a critical study",
  author    = "Rivas, {\'A}ngel and K Plato, A Douglas and Huelga, Susana F and
               B Plenio, Martin",
  journal   = "New J. Phys.",
  publisher = "IOP Publishing",
  volume    =  12,
  number    =  11,
  pages     = "113032",
  month     =  nov,
  year      =  2010,
  url = {https://doi.org/10.1088/1367-2630/12/11/113032}
}

@ARTICLE{Rau1963,
  title     = "Relaxation phenomena in spin and harmonic oscillator systems",
  author    = "Rau, Jayaseetha",
  journal   = "Phys. Rev.",
  publisher = "American Physical Society (APS)",
  volume    =  129,
  number    =  4,
  pages     = "1880--1888",
  month     =  feb,
  year      =  1963,
  copyright = "http://link.aps.org/licenses/aps-default-license",
  url = "https://doi.org/10.1103/PhysRev.129.1880"
}

@ARTICLE{Palma2022,
  title     = "Quantum collision models: Open system dynamics from repeated
               interactions",
  author    = "Ciccarello, Francesco and Lorenzo, Salvatore and Giovannetti,
               Vittorio and Palma, G Massimo",
  journal   = "Phys. Rep.",
  publisher = "Elsevier BV",
  volume    =  954,
  pages     = "1--70",
  month     =  apr,
  year      =  2022,
  url = "https://doi.org/10.1016/j.physrep.2022.01.001"
}

@ARTICLE{Merkli2014,
  title     = "Repeated interactions in open quantum systems",
  author    = "Bruneau, Laurent and Joye, Alain and Merkli, Marco",
  journal   = "J. Math. Phys.",
  publisher = "AIP Publishing",
  volume    =  55,
  number    =  7,
  pages     = "075204",
  month     =  jul,
  year      =  2014,
  url = "https://doi.org/10.1063/1.4879240"
}

@ARTICLE{Pocrnic2025,
  author = {Pocrnic, Matthew and Segal, Dvira and Wiebe, Nathan},
  title = {Quantum simulation of Lindbladian dynamics via repeated interactions},
  journal = {Journal of Physics A: Mathematical and Theoretical},
  volume = {58},
  pages = {305302},
  year = {2025},
  doi = {10.1088/1751-8121/adebc4},
  url = "https://doi.org/10.1088/1751-8121/adebc4"
}

@ARTICLE{Prositto2026,
  title     = "Collisional model with dissipative and dephasing baths:
               nonadditive effects at strong coupling",
  author    = "Prositto, Alessandro and Ramon-Escandell, Carlos and Segal, Dvira",
  journal   = "New J. Phys.",
  publisher = "IOP Publishing",
  volume    =  28,
  number    =  1,
  pages     = "014502",
  month     =  jan,
  year      =  2026,
  copyright = "https://creativecommons.org/licenses/by/4.0/",
  url = "https://doi.org/10.1088/1367-2630/ae2e32"
}

@ARTICLE{Campbell2021,
  title     = "Collision models in open system dynamics: A versatile tool for
               deeper insights?",
  author    = "Campbell, Steve and Vacchini, Bassano",
  journal   = "Europhys. Lett.",
  publisher = "IOP Publishing",
  volume    =  133,
  number    =  6,
  pages     = "60001",
  month     =  mar,
  year      =  2021,
  url = "https://doi.org/10.1209/0295-5075/133/60001"
}

@article{Campbell20,
  author  = {Guarnieri, G. and Morrone, D. and {\c{C}}akmak, B. and Plastina, F. and Campbell, S.},
  title   = {Non-equilibrium steady-states of memoryless quantum collision models},
  journal = {Physics Letters A},
  volume  = {384},
  pages   = {126576},
  year    = {2020},
  doi     = {10.1016/j.physleta.2020.126576},
  url = "https://doi.org/10.1016/j.physleta.2020.126576"
}

@ARTICLE{Buzek2005,
  title     = "Description of quantum dynamics of open systems based on
               collision-like models",
  author    = "Ziman, M{\'a}rio and {\v S}telmachovi{\v c}, Peter and Bu{\v
               z}ek, Vladim{\'\i}r",
  journal   = "Open Syst. Inf. Dyn.",
  publisher = "World Scientific Pub Co Pte Lt",
  volume    =  12,
  number    =  01,
  pages     = "81--91",
  month     =  mar,
  year      =  2005,
  url = "https://link.springer.com/article/10.1007/s11080-005-0488-0"
}

@article{Poletti2023,
  title = {Simulating quantum transport via collisional models on a digital quantum computer},
  author = {Erbanni, Rebecca and Xu, Xiansong and Demarie, Tommaso F. and Poletti, Dario},
  journal = {Phys. Rev. A},
  volume = {108},
  issue = {3},
  pages = {032619},
  numpages = {10},
  year = {2023},
  month = {Sep},
  publisher = {American Physical Society},
  doi = {10.1103/PhysRevA.108.032619},
  url = {https://link.aps.org/doi/10.1103/PhysRevA.108.032619}
}

@ARTICLE{Buzek2012,
  title     = "Simulation of indivisible qubit channels in collision models",
  author    = "Ryb{\'a}r, Tom{\'a}{\v s} and Filippov, Sergey N and Ziman,
               M{\'a}rio and Bu{\v z}ek, Vladim{\'\i}r",
  journal   = "J. Phys. B At. Mol. Opt. Phys.",
  publisher = "IOP Publishing",
  volume    =  45,
  number    =  15,
  pages     = "154006",
  month     =  aug,
  year      =  2012,
  url = "https://doi.org/10.1088/0953-4075/45/15/154006"
}

@ARTICLE{Giovannetti2013,
  title = {Collision-model-based approach to non-Markovian quantum dynamics},
  author = {Ciccarello, F. and Palma, G. M. and Giovannetti, V.},
  journal = {Phys. Rev. A},
  volume = {87},
  issue = {4},
  pages = {040103},
  numpages = {5},
  year = {2013},
  month = {Apr},
  publisher = {American Physical Society},
  doi = {10.1103/PhysRevA.87.040103},
  url = {https://link.aps.org/doi/10.1103/PhysRevA.87.040103}
}

@article{Strunz2016NonMarkovian,
  title = {Collision model for non-Markovian quantum dynamics},
  author = {Kretschmer, Silvan and Luoma, Kimmo and Strunz, Walter T.},
  journal = {Phys. Rev. A},
  volume = {94},
  issue = {1},
  pages = {012106},
  numpages = {9},
  year = {2016},
  month = {Jul},
  publisher = {American Physical Society},
  doi = {10.1103/PhysRevA.94.012106},
  url = {https://link.aps.org/doi/10.1103/PhysRevA.94.012106}
}

@ARTICLE{Mauro2018,
  title     = "Reconciliation of quantum local master equations with
               thermodynamics",
  author    = "De Chiara, Gabriele and Landi, Gabriel and Hewgill, Adam and
               Reid, Brendan and Ferraro, Alessandro and Roncaglia, Augusto J
               and Antezza, Mauro",
  journal   = "New J. Phys.",
  publisher = "IOP Publishing",
  volume    =  20,
  number    =  11,
  pages     = "113024",
  month     =  nov,
  year      =  2018,
  copyright = "http://creativecommons.org/licenses/by/3.0/",
  url = "https://doi.org/10.1088/1367-2630/aaecee"
}

@ARTICLE{Xia2022,
  title     = "Features of quantum thermodynamics induced by common
               environments based on collision model",
  author    = "Huang, Rui and Man, Zhong-Xiao and Zhang, Ying-Jie and Xia, Yun-Jie",
  journal   = "EPJ Quantum Technol.",
  publisher = "Springer Science and Business Media LLC",
  volume    =  9,
  pages = 28,
  number    =  1,
  month     =  dec,
  year      =  2022,
  copyright = "https://creativecommons.org/licenses/by/4.0",
  url = "https://doi.org/10.1140/epjqt/s40507-022-00148-9"
}

@article{Brandes2017,
  title = {Quantum and Information Thermodynamics: A Unifying Framework Based on Repeated Interactions},
  author = {Strasberg, Philipp and Schaller, Gernot and Brandes, Tobias and Esposito, Massimiliano},
  journal = {Phys. Rev. X},
  volume = {7},
  issue = {2},
  pages = {021003},
  numpages = {33},
  year = {2017},
  month = {Apr},
  publisher = {American Physical Society},
  doi = {10.1103/PhysRevX.7.021003},
  url = {https://link.aps.org/doi/10.1103/PhysRevX.7.021003}
}

@article{Landi2019Thermometry,
  title = {Collisional Quantum Thermometry},
  author = {Seah, Stella and Nimmrichter, Stefan and Grimmer, Daniel and Santos, Jader P. and Scarani, Valerio and Landi, Gabriel T.},
  journal = {Phys. Rev. Lett.},
  volume = {123},
  issue = {18},
  pages = {180602},
  numpages = {6},
  year = {2019},
  month = {Oct},
  publisher = {American Physical Society},
  doi = {10.1103/PhysRevLett.123.180602},
  url = {https://link.aps.org/doi/10.1103/PhysRevLett.123.180602}
}

@ARTICLE{Zambrini2022,
  title     = "Quantum thermodynamics under continuous monitoring: A general
               framework",
  author    = "Manzano, Gonzalo and Zambrini, Roberta",
  journal   = "AVS Quantum Sci.",
  publisher = "American Vacuum Society",
  volume    =  4,
  number    =  2,
  pages     = "025302",
  month     =  jun,
  year      =  2022,
  copyright = "https://creativecommons.org/licenses/by/4.0/",
  url = "https://doi.org/10.1116/5.0079886"
}

@article{Strasberg2019,
  title = {Repeated Interactions and Quantum Stochastic Thermodynamics at Strong Coupling},
  author = {Strasberg, Philipp},
  journal = {Phys. Rev. Lett.},
  volume = {123},
  issue = {18},
  pages = {180604},
  numpages = {7},
  year = {2019},
  month = {Oct},
  publisher = {American Physical Society},
  doi = {10.1103/PhysRevLett.123.180604},
  url = {https://link.aps.org/doi/10.1103/PhysRevLett.123.180604}
}

@ARTICLE{Gerasimov2025,
  title     = "Repeated temperature measurements in quantum thermodynamics",
  author    = "Gerasimov, N M and Teretenkov, A E",
  journal   = "Lobachevskii J. Math.",
  publisher = "Pleiades Publishing Ltd",
  volume    =  46,
  number    =  6,
  pages     = "2501--2512",
  month     =  jun,
  year      =  2025,
  copyright = "https://www.springernature.com/gp/researchers/text-and-data-mining",
  url = "https://link.springer.com/article/10.1134/S1995080225608082"
}

@ARTICLE{Pautrat2017,
  title     = "Landauer's principle in repeated interaction systems",
  author    = "Hanson, Eric P and Joye, Alain and Pautrat, Yan and
               Raqu{\'e}pas, Renaud",
  journal   = "Commun. Math. Phys.",
  publisher = "Springer Science and Business Media LLC",
  volume    =  349,
  number    =  1,
  pages     = "285--327",
  month     =  jan,
  year      =  2017,
  url = "https://link.springer.com/article/10.1007/s00220-016-2751-3"
}

@ARTICLE{Kurizki2008,
  title     = "Thermodynamic control by frequent quantum measurements",
  author    = "Erez, Noam and Gordon, Goren and Nest, Mathias and Kurizki, Gershon",
  journal   = "Nature",
  publisher = "Springer Science and Business Media LLC",
  volume    =  452,
  number    =  7188,
  pages     = "724--727",
  month     =  apr,
  year      =  2008,
  url = "https://doi.org/10.1038/nature06873"
}

@article{Kurizki2019,
  title = {Collectively enhanced thermalization via multiqubit collisions},
  author = "Manatuly, Angsar and Niedenzu, Wolfgang and Rom{\'a}n-Ancheyta,
               Ricardo and {\c C}akmak, Bar{\i}{\c s} and M{\"u}stecapl{\i}o{\u
               g}lu, {\"O}zg{\"u}r E and Kurizki, Gershon",
  journal = {Phys. Rev. E},
  volume = {99},
  issue = {4},
  pages = {042145},
  numpages = {9},
  year = {2019},
  month = {Apr},
  publisher = {American Physical Society},
  doi = {10.1103/PhysRevE.99.042145},
  url = {https://link.aps.org/doi/10.1103/PhysRevE.99.042145}
}

@ARTICLE{DeChiara2022,
  title     = "Exploiting coherence for quantum thermodynamic advantage",
  author    = "Hammam, Kenza and Leitch, Heather and Hassouni, Yassine and De Chiara, Gabriele",
  journal   = "New J. Phys.",
  publisher = "IOP Publishing",
  volume    =  24,
  number    =  11,
  pages     = "113053",
  month     =  nov,
  year      =  2022,
  copyright = "http://creativecommons.org/licenses/by/4.0",
  url = "https://doi.org/10.1088/1367-2630/aca49b"
}

@article{Landi2019,
  title = {Thermodynamics of Weakly Coherent Collisional Models},
  author = {Rodrigues, Franklin L. S. and De Chiara, Gabriele and Paternostro, Mauro and Landi, Gabriel T.},
  journal = {Phys. Rev. Lett.},
  volume = {123},
  issue = {14},
  pages = {140601},
  numpages = {6},
  year = {2019},
  month = {Oct},
  publisher = {American Physical Society},
  doi = {10.1103/PhysRevLett.123.140601},
  url = {https://link.aps.org/doi/10.1103/PhysRevLett.123.140601}
}

@ARTICLE{Parrondo2022,
  title     = "Quantum collisional thermostats",
  author    = "Tabanera, Jorge and Luque, In{\'e}s and Jacob, Samuel L and
               Esposito, Massimiliano and Barra, Felipe and Parrondo, Juan M R",
  journal   = "New J. Phys.",
  publisher = "IOP Publishing",
  volume    =  24,
  number    =  2,
  pages     = "023018",
  month     =  feb,
  year      =  2022,
  copyright = "https://creativecommons.org/licenses/by/4.0/",
  url = "https://doi.org/10.1088/1367-2630/ac4923"
}

@article{Breuer2022,
  title = {Open-system approach to nonequilibrium quantum thermodynamics at arbitrary coupling},
  author = {Colla, Alessandra and Breuer, Heinz-Peter},
  journal = {Phys. Rev. A},
  volume = {105},
  issue = {5},
  pages = {052216},
  numpages = {8},
  year = {2022},
  month = {May},
  publisher = {American Physical Society},
  doi = {10.1103/PhysRevA.105.052216},
  url = {https://link.aps.org/doi/10.1103/PhysRevA.105.052216}
}

@ARTICLE{Kosloff2014Machine,
  title     = "The multilevel four-stroke swap engine and its environment",
  author    = "Uzdin, Raam and Kosloff, Ronnie",
  journal   = "New J. Phys.",
  publisher = "IOP Publishing",
  volume    =  16,
  number    =  9,
  pages     = "095003",
  month     =  sep,
  year      =  2014,
  copyright = "http://creativecommons.org/licenses/by/3.0/",
  url = "http://dx.doi.org/10.1088/1367-2630/16/9/095003"
}

@article{Juan2019,
  title = {Autonomous thermal machine for amplification and control of energetic coherence},
  author = {Manzano, Gonzalo and Silva, Ralph and Parrondo, Juan M. R.},
  journal = {Phys. Rev. E},
  volume = {99},
  issue = {4},
  pages = {042135},
  numpages = {14},
  year = {2019},
  month = {Apr},
  publisher = {American Physical Society},
  doi = {10.1103/PhysRevE.99.042135},
  url = {https://link.aps.org/doi/10.1103/PhysRevE.99.042135}
}

@article{DeChiara2020,
  title = {Three-qubit refrigerator with two-body interactions},
  author = {Hewgill, Adam and Gonz\'alez, J. Onam and Palao, Jos\'e P. and Alonso, Daniel and Ferraro, Alessandro and De Chiara, Gabriele},
  journal = {Phys. Rev. E},
  volume = {101},
  issue = {1},
  pages = {012109},
  numpages = {10},
  year = {2020},
  month = {Jan},
  publisher = {American Physical Society},
  doi = {10.1103/PhysRevE.101.012109},
  url = {https://link.aps.org/doi/10.1103/PhysRevE.101.012109}
}

@article{Bellomo2021,
  title = {Power maximization of two-stroke quantum thermal machines},
  author = {Piccione, Nicol\`o and De Chiara, Gabriele and Bellomo, Bruno},
  journal = {Phys. Rev. A},
  volume = {103},
  issue = {3},
  pages = {032211},
  numpages = {15},
  year = {2021},
  month = {Mar},
  publisher = {American Physical Society},
  doi = {10.1103/PhysRevA.103.032211},
  url = {https://link.aps.org/doi/10.1103/PhysRevA.103.032211}
}

@article{Landi2020,
  title = {Stroboscopic two-stroke quantum heat engines},
  author = {Molitor, Otavio A. D. and Landi, Gabriel T.},
  journal = {Phys. Rev. A},
  volume = {102},
  issue = {4},
  pages = {042217},
  numpages = {10},
  year = {2020},
  month = {Oct},
  publisher = {American Physical Society},
  doi = {10.1103/PhysRevA.102.042217},
  url = {https://link.aps.org/doi/10.1103/PhysRevA.102.042217}
}

@article{Mauro2020,
  title = {Quantum machines powered by correlated baths},
  author = {De Chiara, Gabriele and Antezza, Mauro},
  journal = {Phys. Rev. Res.},
  volume = {2},
  issue = {3},
  pages = {033315},
  numpages = {9},
  year = {2020},
  month = {Aug},
  publisher = {American Physical Society},
  doi = {10.1103/PhysRevResearch.2.033315},
  url = {https://link.aps.org/doi/10.1103/PhysRevResearch.2.033315}
}

@ARTICLE{Xia2022ThermalMachines,
  title     = "Work costs and operating regimes for different manners of
               system-reservoir interactions via collision model",
  author    = "Wang, Ying and Man, Zhong-Xiao and Zhang, Ying-Jie and Xia, Yun-Jie",
  journal   = "New J. Phys.",
  publisher = "IOP Publishing",
  volume    =  24,
  number    =  5,
  pages     = "053030",
  month     =  may,
  year      =  2022,
  copyright = "https://creativecommons.org/licenses/by/4.0/",
  url = "https://doi.org/10.1088/1367-2630/ac6a01"
}

@ARTICLE{Wehner2019,
  title     = "The maximum efficiency of nano heat engines depends on more than
               temperature",
  author    = "Woods, Mischa P and Ng, Nelly Huei Ying and Wehner, Stephanie",
  journal   = "Quantum",
  publisher = "Verein zur Forderung des Open Access Publizierens in den
               Quantenwissenschaften",
  volume    =  3,
  number    =  177,
  pages     = "177",
  month     =  aug,
  year      =  2019,
  url = "https://doi.org/10.22331/q-2019-08-19-177"
}

@article{Goold2023ThermalMachines,
  title = {Thermodynamics of a continuously monitored double-quantum-dot heat engine in the repeated interactions framework},
  author = {Bettmann, Laetitia P. and Kewming, Michael J. and Goold, John},
  journal = {Phys. Rev. E},
  volume = {107},
  issue = {4},
  pages = {044102},
  numpages = {12},
  year = {2023},
  month = {Apr},
  publisher = {American Physical Society},
  doi = {10.1103/PhysRevE.107.044102},
  url = {https://link.aps.org/doi/10.1103/PhysRevE.107.044102}
}

@ARTICLE{Widera2021,
  title     = "A quantum heat engine driven by atomic collisions",
  author    = "Bouton, Quentin and Nettersheim, Jens and Burgardt, Sabrina and
               Adam, Daniel and Lutz, Eric and Widera, Artur",
  journal   = "Nat. Commun.",
  publisher = "Springer Science and Business Media LLC",
  volume    =  12,
  number    =  1,
  pages     = "2063",
  month     =  apr,
  year      =  2021,
  copyright = "https://creativecommons.org/licenses/by/4.0",
  url = "https://doi.org/10.1038/s41467-021-22222-z"
}

@ARTICLE{Scully2003,
  title     = "Extracting work from a single heat bath via vanishing quantum
               coherence",
  author    = "Scully, Marlan O and Zubairy, M Suhail and Agarwal, Girish S and
               Walther, Herbert",
  journal   = "Science",
  publisher = "American Association for the Advancement of Science (AAAS)",
  volume    =  299,
  number    =  5608,
  pages     = "862--864",
  month     =  feb,
  year      =  2003,
  url = "https://doi.org/10.1126/science.1078955"
}

@ARTICLE{Petruccione2019,
  title     = "Quantum coherence, many-body correlations, and non-thermal
               effects for autonomous thermal machines",
  author    = "Latune, C L and Sinayskiy, I and Petruccione, F",
  journal   = "Sci. Rep.",
  publisher = "Springer Science and Business Media LLC",
  volume    =  9,
  number    =  1,
  pages     = "3191",
  month     =  feb,
  year      =  2019,
  copyright = "https://creativecommons.org/licenses/by/4.0",
  url = "https://doi.org/10.1038/s41598-019-39300-4"
}

@ARTICLE{Kurziki2018QuantumAdvantage,
  title     = "Cooperative many-body enhancement of quantum thermal machine
               power",
  author    = "Niedenzu, Wolfgang and Kurizki, Gershon",
  journal   = "New J. Phys.",
  publisher = "IOP Publishing",
  volume    =  20,
  number    =  11,
  pages     = "113038",
  month     =  nov,
  year      =  2018,
  copyright = "http://creativecommons.org/licenses/by/3.0/",
  url = "https://doi.org/10.1088/1367-2630/aaed55"
}

@ARTICLE{Manzano2021,
  title     = "Optimizing autonomous thermal machines powered by energetic
               coherence",
  author    = "Hammam, Kenza and Hassouni, Yassine and Fazio, Rosario and
               Manzano, Gonzalo",
  journal   = "New J. Phys.",
  publisher = "IOP Publishing",
  volume    =  23,
  number    =  4,
  pages     = "043024",
  month     =  apr,
  year      =  2021,
  copyright = "https://creativecommons.org/licenses/by/4.0/",
  url = "https://doi.org/10.1088/1367-2630/abeb47"
}

@article{Poem2019,
  title = {Experimental Demonstration of Quantum Effects in the Operation of Microscopic Heat Engines},
  author = {Klatzow, James and Becker, Jonas N. and Ledingham, Patrick M. and Weinzetl, Christian and Kaczmarek, Krzysztof T. and Saunders, Dylan J. and Nunn, Joshua and Walmsley, Ian A. and Uzdin, Raam and Poem, Eilon},
  journal = {Phys. Rev. Lett.},
  volume = {122},
  issue = {11},
  pages = {110601},
  numpages = {6},
  year = {2019},
  month = {Mar},
  publisher = {American Physical Society},
  doi = {10.1103/PhysRevLett.122.110601},
  url = {https://link.aps.org/doi/10.1103/PhysRevLett.122.110601}
}

@ARTICLE{Manzano2025,
  title     = "Certifying quantum enhancements in thermal machines beyond the
               Thermodynamic Uncertainty Relation",
  author    = "Almanza-Marrero, Jos{\'e} A and Manzano, Gonzalo",
  journal = "Quantum",
  volume    =  9,
  number    =  1878,
  pages     = "1878",
  month     =  oct,
  year      =  2025,
  url = "https://doi.org/10.22331/q-2025-10-07-1878"
}

@ARTICLE{Brunner2017,
  title     = "Markovian master equations for quantum thermal machines: local
               versus global approach",
  author    = "Hofer, Patrick P and Perarnau-Llobet, Mart{\'\i} and Miranda, L
               David M and Haack, G{\'e}raldine and Silva, Ralph and Brask,
               Jonatan Bohr and Brunner, Nicolas",
  journal   = "New J. Phys.",
  publisher = "IOP Publishing",
  volume    =  19,
  number    =  12,
  pages     = "123037",
  month     =  dec,
  year      =  2017,
  copyright = "http://creativecommons.org/licenses/by/3.0/",
  url = "https://doi.org/10.1088/1367-2630/aa964f"
}

@ARTICLE{Cleverson2025,
  title     = "Innovative designs and insights into quantum thermal machines",
  author    = "L{\'u}cio, Aline Duarte and Rojas, Moises and Filgueiras,
               Cleverson",
  journal   = "Quantum Rep.",
  publisher = "MDPI AG",
  volume    =  7,
  number    =  2,
  pages     = "26",
  month     =  jun,
  year      =  2025,
  copyright = "https://creativecommons.org/licenses/by/4.0/",
  url = "https://doi.org/10.3390/quantum7020026"
}

@ARTICLE{Karen2025,
  title     = "Quantum many-body thermal machines enabled by atom-atom
               correlations",
  author    = "Watson, Raymon S and Kheruntsyan, Karen",
  journal   = "SciPost Phys.",
  publisher = "Stichting SciPost",
  volume    =  18,
  pages = 190,
  number    =  6,
  month     =  jun,
  year      =  2025,
  copyright = "https://creativecommons.org/licenses/by/4.0",
  url = "https://doi.org/10.21468/SciPostPhys.18.6.190"
}

@article{Lutz2014,
  title = {Nanoscale Heat Engine Beyond the {C}arnot Limit},
  author = {Ro\ss{}nagel, J. and Abah, O. and Schmidt-Kaler, F. and Singer, K. and Lutz, E.},
  journal = {Phys. Rev. Lett.},
  volume = {112},
  issue = {3},
  pages = {030602},
  numpages = {5},
  year = {2014},
  month = {Jan},
  publisher = {American Physical Society},
  doi = {10.1103/PhysRevLett.112.030602},
  url = {https://link.aps.org/doi/10.1103/PhysRevLett.112.030602}
}

@article{Manzano2016,
  title = {Entropy production and thermodynamic power of the squeezed thermal reservoir},
  author = {Manzano, Gonzalo and Galve, Fernando and Zambrini, Roberta and Parrondo, Juan M. R.},
  journal = {Phys. Rev. E},
  volume = {93},
  issue = {5},
  pages = {052120},
  numpages = {10},
  year = {2016},
  month = {May},
  publisher = {American Physical Society},
  doi = {10.1103/PhysRevE.93.052120},
  url = {https://link.aps.org/doi/10.1103/PhysRevE.93.052120}
}

@article{Yi2012,
  title = {Effects of reservoir squeezing on quantum systems and work extraction},
  author = {Huang, X. L. and Wang, Tao and Yi, X. X.},
  journal = {Phys. Rev. E},
  volume = {86},
  issue = {5},
  pages = {051105},
  numpages = {6},
  year = {2012},
  month = {Nov},
  publisher = {American Physical Society},
  doi = {10.1103/PhysRevE.86.051105},
  url = {https://link.aps.org/doi/10.1103/PhysRevE.86.051105}
}

@ARTICLE{DelCampo2016,
  title     = "Quantum supremacy of many-particle thermal machines",
  author    = "Jaramillo, J and Beau, M and Campo, A del",
  journal   = "New J. Phys.",
  publisher = "IOP Publishing",
  volume    =  18,
  number    =  7,
  pages     = "075019",
  month     =  jul,
  year      =  2016,
  copyright = "http://creativecommons.org/licenses/by/3.0/",
  url  = "http://dx.doi.org/10.1088/1367-2630/18/7/075019"
}

@ARTICLE{Ghosh2023,
  title     = "Quantum advantage of thermal machines with Bose and Fermi gases",
  author    = "Sur, Saikat and Ghosh, Arnab",
  journal   = "Entropy (Basel)",
  publisher = "MDPI AG",
  volume    =  25,
  number    =  2,
  pages     = "372",
  month     =  feb,
  year      =  2023,
  copyright = "https://creativecommons.org/licenses/by/4.0/",
  url = "https://doi.org/10.3390/e25020372"
}

@ARTICLE{Divakaran2021,
  title     = "Many-body quantum thermal machines",
  author    = "Mukherjee, Victor and Divakaran, Uma",
  journal   = "J. Phys. Condens. Matter",
  publisher = "IOP Publishing",
  volume    =  33,
  number    =  45,
  pages     = "454001",
  month     =  aug,
  year      =  2021,
  copyright = "https://publishingsupport.iopscience.iop.org/iop-standard/v1",
  url = {https://doi.org/10.1088/1361-648X/ac1b60}
}

\end{document}